\documentclass[apj]{emulateapj}
\usepackage{epstopdf}
\usepackage{longtable}
\usepackage{fancyvrb}
\usepackage{multirow}
\usepackage{verbatim}
\usepackage{longtable}
\usepackage[caption=false]{subfig}
\usepackage{float}
\usepackage{color}
\newcommand{\wfcthree}{WFC3}
\newcommand{\Hb}{$H$}
\newcommand{\Ib}{$I$}
\newcommand{\Bb}{$B$}
\newcommand{\sdssg}{$g$}
\newcommand{\sdssr}{$r$}

\newcommand{\sdssz}{$z$}
\newcommand{\HbFWHM}{$0\farcs151$}

\newcommand{\lampeak}{$\lambda_{\rm{peak}}$}
\newcommand{\Hlampeak}{$1.545$ \micron}
\newcommand{\Ilampeak}{$8353$ \AA}
\newcommand{\Blampeak}{$4320$ \AA}
\newcommand{\BIcolor}{$B-I$}
\newcommand{\IHcolor}{$I-H$}
\newcommand{\BHcolor}{$B-H$}

\newcommand{\hstname}{\emph{Hubble Space Telescope}}
\newcommand{\hst}{\emph{HST}}
\newcommand{\ch}{\emph{Chandra}}
\newcommand{\lynx}{\emph{Lynx}}
\newcommand{\sdssname}{Sloan Digital Sky Survey}
\newcommand{\sdss}{SDSS}

\newcommand{\wisename}{\emph{Wide-field Infrared Survey Explorer}}
\newcommand{\swift}{\emph{Swift}}
\newcommand{\batname}{Burst Alert Telescope}
\newcommand{\bat}{BAT}

\newcommand{\Mone}{$M_{1}$}
\newcommand{\Mtwo}{$M_{2}$}
\newcommand{\mratio}{$M_{1}/M_{2}$}

\newcommand{\deltas}{$\Delta S$}

\newcommand{\Rv}{Rv}

\newcommand{\mpa}{MPA-JHU}

\newcommand{\prim}{primary}
\newcommand{\second}{secondary}
\newcommand{\fitdim}{25}

\newcommand{\asymm}{$A$}

\newcommand{\Mgal}{$M_{\rm{Gal.}}$}

\newcommand{\RAX}{RA$_{\rm{X-ray}}$}
\newcommand{\DECX}{DEC$_{\rm{X-ray}}$}

\newcommand{\deltaANGPri}{$\Delta \Theta_{1}$}
\newcommand{\deltaANGSec}{$\Delta \Theta_{2}$}
\newcommand{\deltaSPri}{$\Delta \rm{S}_{1}$}
\newcommand{\deltaSSec}{$\Delta \rm{S}_{2}$}
\newcommand{\PAPri}{PA$_{1}$}
\newcommand{\PASec}{PA$_{2}$}

\newcommand{\uLum}{erg s$^{-1}$}
\newcommand{\Msun}{$M_{\odot}$}
\newcommand{\uKPC}{kpc}
\newcommand{\uDEG}{$^{\circ}$}
\newcommand{\uARCSEC}{$''$}
\newcommand{\uSEC}{s}
\newcommand{\uHMS}{hh:mm:ss.sss}
\newcommand{\uDMS}{dd:mm:ss.ss}
\newcommand{\uNH}{cm$^{-2}$}
\newcommand{\uCNTS}{counts}
\newcommand{\uSIG}{$\sigma$}

\newcommand{\nhexgal}{n$_{H}$}

\newcommand{\LXhard}{$L_{\rm{2-10keV}}$}

\newcommand{\XScnts}{$S$}
\newcommand{\XHcnts}{$H$}
\newcommand{\Xcnts}{$H+S$}
\newcommand{\Xdetsig}{Det. Sig.}

\newcommand{\MBH}{$M_{BH}$}
\newcommand{\LBol}{$L_{\rm{Bol.}}$}

\newcommand{\BIonecolor}{$B_{1}-I_{1}$}
\newcommand{\IHonecolor}{$I_{1}-H_{1}$}
\newcommand{\BHonecolor}{$B_{1}-H_{1}$}
\newcommand{\BItwocolor}{$B_{2}-I_{2}$}
\newcommand{\IHtwocolor}{$I_{2}-H_{2}$}
\newcommand{\BHtwocolor}{$B_{2}-H_{2}$}

\newcommand{\oiii}{[\ion{O}{3}]$\lambda5007$}

\newcommand{\niib}{[\ion{N}{2}]$\lambda6583$}

\newcommand{\hb}{H$\beta$}
\newcommand{\ha}{H$\alpha$}

\newcommand{\typeII}{Type 2}

\newcommand{\ciao}{\texttt{CIAO}}
\newcommand{\sherpa}{\texttt{Sherpa}}

\newcommand{\gf}{\textsc{Galfit}}
\newcommand{\se}{\texttt{Source Extractor}}

\newcommand{\TARGname}{Target Name}
\newcommand{\name}{Name}

\newcommand{\offone}{J0813$+$5418}

\newcommand{\offtwo}{J0940$+$3113}

\newcommand{\offthree}{J1021$+$1305}

\newcommand{\offfour}{J1114$+$4036}

\newcommand{\offfive}{J1234$+$4751}

\newcommand{\offsix}{J2125$-$0713}

\newcommand{\offAGNhstsztext}{six}
\newcommand{\offAGNsdsssztext}{three}
\newcommand{\offAGNplotsztext}{six} 
\newcommand{\dualAGNCsztext}{six}
\newcommand{\dualAGNLsztext}{three}
\newcommand{\dualAGNtotsztext}{nine} 
\newcommand{\dualAGNplotsztext}{ten} 

\newcommand{\paperI}{Paper I}

\newcommand{\apsizefactext}{three}

\newcommand{\oneHtwoMgal}{$\sim 2\times 10^{6}$} 
\newcommand{\oneHtwoMBH}{$\sim 8\times 10^{3}$} 

\newcommand{\twophysep}{$6.6$}

\newcommand{\threeHtwoMgal}{$\sim 3\times 10^{8}$} 
\newcommand{\threeHtwoMBH}{$\sim 8\times 10^{5}$} 

\newcommand{\fourphysep}{$1.2$}

\newcommand{\fiveHtwoMgal}{$5.1\times 10^{10}$}
\newcommand{\fiveHtwoMBH}{$1.2\times 10^{8}$}

\newcommand{\fivemratio}{$2$}

\newcommand{\sixHtwoMgal}{$1.1\times 10^{8}$}
\newcommand{\sixHtwoMBH}{$3.6\times 10^{5}$}

\newcommand{\sixmratio}{$1000$}

\newcommand{\MRATIOthresh}{$ 4$}

\newcommand{\Offnhmean}{$6.3\times 10^{20}$}
\newcommand{\Dualnhmean}{$8.6\times 10^{21}$}

\newcommand{\OffMRATIOmean}{$33.8$}

\newcommand{\DualMRATIOmean}{$4.6$}

\newcommand{\OffDualMRATIOmeandiffsig}{$2.0\sigma$}
\newcommand{\OffDualMRATIOKSstat}{$0.6$}
\newcommand{\OffDualMRATIOKSprob}{$4.7\times 10^{-2}$}

\newcommand{\OffDualLBolLOIIIMRATIOfracsig}{$3.4\sigma$}

\newcommand{\MAJORFRACZAVGLOTZ}{$0.1$}
\newcommand{\MINORFRACZAVGLOTZ}{$0.9$}

\newcommand{\OffDualASYMMKSTstat}{$0.4$}
\newcommand{\OffDualASYMMKSTprob}{$0.5$}

\newcommand{\OffexcMAJORNUMfour}{$ 0$}

\newcommand{\OffexcMINORNUMfour}{$ 5$}

\newcommand{\xraysigdetone}{$3.7$}
\newcommand{\xraysigdettwo}{$2.3$}
\newcommand{\xraysigdetthree}{$2.3$}
\newcommand{\xraysigdetfourNE}{$3.8$}
\newcommand{\xraysigdetfourSW}{$3.3$}
\newcommand{\xraysigdetfive}{$3.6$}
\newcommand{\xraysigdetsix}{$16.1$}

\newcommand{\xrayScntsone}{$10.5_{-3.4}^{+2.9}$}
\newcommand{\xrayHcntsone}{$6.6_{-2.8}^{+2.0}$}
\newcommand{\xraycntsone}{$17.5_{-4.5}^{+3.5}$}
\newcommand{\xrayScntstwo}{$3.6_{-2.1}^{+1.5}$}
\newcommand{\xrayHcntstwo}{$2.5_{-1.8}^{+1.2}$}
\newcommand{\xraycntstwo}{$6.6_{-3.1}^{+1.8}$}
\newcommand{\xrayScntsthree}{$3.6_{-2.1}^{+1.4}$}
\newcommand{\xrayHcntsthree}{$1.7_{-1.4}^{+0.8}$}
\newcommand{\xraycntsthree}{$5.6_{-2.8}^{+1.6}$}
\newcommand{\xrayScntsfourNE}{$8.6_{-3.1}^{+2.4}$}
\newcommand{\xrayHcntsfourNE}{$7.6_{-3.0}^{+2.3}$}
\newcommand{\xraycntsfourNE}{$16.8_{-4.6}^{+3.2}$}
\newcommand{\xrayScntsfourSW}{$4.7_{-2.3}^{+1.8}$}
\newcommand{\xrayHcntsfourSW}{$6.6_{-2.7}^{+2.2}$}
\newcommand{\xraycntsfourSW}{$11.8_{-3.2}^{+4.0}$}
\newcommand{\xrayScntsfive}{$0.7_{-0.7}^{+1.2}$}
\newcommand{\xrayHcntsfive}{$11.6_{-3.6}^{+3.0}$}
\newcommand{\xraycntsfive}{$12.7_{-4.1}^{+2.7}$}
\newcommand{\xrayScntssix}{$204.6_{-15.1}^{+13.6}$}
\newcommand{\xrayHcntssix}{$84.5_{-9.6}^{+8.5}$}
\newcommand{\xraycntssix}{$289.6_{-17.4}^{+16.4}$}

\shorttitle{Spatially Offset AGN III}
\shortauthors{Barrows et al.}

\begin{document}

\submitted{Accepted for publication in ApJ}

\title{Spatially Offset Active Galactic Nuclei III: Discovery of Late-Stage Galaxy Mergers with \emph{The Hubble Space Telescope}}

\author{R. Scott Barrows$^{1}$, Julia M. Comerford$^{1}$, and Jenny E. Greene$^{2}$}
\affiliation{$^{1}$Department of Astrophysical and Planetary Sciences, University of Colorado Boulder, Boulder, CO 80309, USA; Robert.Barrows@Colorado.edu}
\address{$^{2}$Department of Astrophysical Sciences, Princeton University, Princeton, NJ 08544, USA}

\bibliographystyle{apj}

\begin{abstract}

Galaxy pairs with separations of only a few kpc represent important stages in the merger-driven growth of supermassive black holes (SMBHs). However, such mergers are difficult to identify observationally due to the correspondingly small angular scales. In \paperI~we presented a method of finding candidate kpc-scale galaxy mergers that is leveraged on the selection of X-ray sources spatially offset from the centers of host galaxies. In this paper we analyze new \emph{Hubble Space Telescope} (\hst) WFC3 imaging for \offAGNhstsztext~of these sources to search for signatures of galaxy mergers. The \hst~imaging reveals that four of the \offAGNhstsztext~systems are on-going galaxy mergers with separations of \fourphysep$-$\twophysep~kpc (offset AGN). The nature of the remaining two spatially offset X-ray sources is ambiguous and may be associated with super-Eddington accretion in X-ray binaries. The ability of this sample to probe small galaxy separations and minor mergers makes it uniquely suited for testing the role of galaxy mergers for AGN triggering. We find that galaxy mergers with only one AGN are predominantly minor mergers with mass ratios similar to the overall population of galaxy mergers.  By comparison, galaxy mergers with two AGN are biased toward major mergers and larger nuclear gas masses. Finally, we find that the level of SMBH accretion increases toward smaller mass ratios (major mergers). This result suggests the mass ratio effects not only the frequency of AGN triggering but also the rate of SMBH growth in mergers.

\end{abstract}


\section{Introduction}
\label{sec:intro}

The growth of supermassive black holes (SMBHs) in galaxy nuclei occurs primarily via the accretion of baryonic matter from the interstellar medium (ISM). During this process, a fraction of the energy generated in accretion disks is dissipated in the form of electromagnetic radiation and results in the observational phenomenon of active galactic nuclei (AGN). A long history of numerical simulations predicts that mergers between galaxies are an effective means of transporting mass into accretion disks and triggering AGN \citep{Hernquist:1989,Mihos:1996,DiMatteo:2005,Springel:2005a,Hopkins2008,Capelo:2015}. 

Indeed, observations of AGN in on-going galaxy mergers support the hypothesis of merger-driven SMBH growth \citep{Sanders:1988,Sanders:1988a,Canalizo:2001}. However, the relevance of mergers among the overall AGN population is unclear \citep{Georgakakis:2009,Kocevski:2012,Simmons:2012,Villforth:2014,Mechtley:2015,Villforth:2016}. Hence, uniform and reliable samples of galaxy mergers are necessary to test this hypothesis. Moreover, fully understanding how AGN evolve throughout mergers requires the ability to trace the SMBHs down to small pair separations and to identify which are actively accreting. 

At large physical separations, most AGN selection methods can suffice. For example, when the galaxy pair separation is larger than the fiber collision limit \citep{Blanton:2001}, optical emission line diagnostics \citep{Baldwin1981,Kewley:2006} from the \sdssname~(\sdss) have provided AGN identifications for individual galaxies in many pair samples \citep{Ellison:2008,Ellison:2011,Scudder:2012,Patton:2013}. More recently, infrared selections of AGN from the \wisename~\citep{Wright:2010} based on magnitude dependent color cuts \citep{Stern:2012,Assef:2013} can be used to associate luminous AGN with galaxies in pairs resolvable by the $6\farcs5$ resolution limit. Unfortunately, these spatial limitations effectively reject most advanced mergers. However, theory predicts that the efficiency of AGN triggering in mergers peaks at separations of $<10$ kpc \citep{Van_Wassenhove:2012,Blecha:2013,Capelo:2015,Steinborn:2016}. While radio observations can probe these merger phases, only $\sim10\%$ of AGN are radio loud and deep observations are required to detect others \citep{Mueller-Sanchez:2015}.

To overcome this observational hurdle, we previously developed a procedure for identifying candidate late-stage galaxy mergers hosting a single AGN (offset AGN) based on spatially offset X-ray sources \citep[][hereafter \paperI]{Barrows:2016}. Utilizing the spatial resolution of the \ch~X-ray Observatory, we constrained the positions of X-ray sources within galaxies so that they can be detected as offset from galactic nuclei and potentially from other discrete X-ray sources within the same galaxy. Detection of a spatially offset AGN signifies a galaxy merger, and in \paperI~we posited that spatially offset X-ray sources may represent kpc-scale galaxy mergers. Due to the small physical separations probed by offset AGN, they are ideally suited for studying the specific conditions under which galaxy mergers can drive SMBH growth. For example, we recently used this sample to track, for the first time, evolution of the AGN merger fraction below separations of $1$ kpc \citep{Barrows:2017}. This result echoes and extends that from samples of larger separation pairs \citep{Ellison:2008,Ellison:2011,Koss:2012,Satyapal:2014} and implies that the probability of observing AGN in mergers is higher under the conditions encountered at late merger stages when the nuclei are heavily enshrouded by gas and dust.

However, to form a comprehensive picture of SMBH growth in mergers, one must understand the physical mechanisms within the merging galaxies that actually drive accretion onto the SMBHs at small separations. Doing so requires testing if properties of the galaxies themselves are connected with SMBH growth, and if they affect the accretion rates or simply increase the probability of AGN triggering. In particular, some studies have hinted that the strongest enhancements in merger-driven AGN triggering occur among galaxy mergers where mass ratios between the more and less massive galaxy are close to unity (major mergers) and most efficiently drive ISM material to the galaxy nuclei \citep{Ellison:2008,comerford:2015,Barrows:2017}. Moreover, numerical simulations predict that galaxy mergers with only a single AGN (offset AGN) preferentially have large mass ratios between the more and less massive galaxy (minor mergers) while galaxy mergers with two AGN (dual AGN) are preferentially associated with major mergers \citep{Steinborn:2016}. 

Simulations also predict that increases in the galaxies' overall supply of gas correspond to a higher probability of AGN triggering \citep{Capelo:2015} and potentially a distinction between offset AGN and dual AGN \citep{Steinborn:2016,Rosas-Guevara:2018}. The supply of nuclear material for accretion may also depend on the location of each SMBH within the merging system itself. However, theory has yet to converge on a clear picture, with theoretical work suggesting that the more luminous AGN will most frequently be associated with the more massive galaxy \citep{Yu:2011} or the less massive galaxy \citep{Capelo:2015}, and that it may depend on the total gas mass \citep{Steinborn:2016}. Overall, these predictions suggest that the galaxy masses and merger morphologies may play an important role in the triggering of AGN in mergers.

To confirm the merger scenario for host galaxies of spatially offset X-ray sources and to test the above predictions for how merger-driven AGN triggering is linked to host galaxy properties in small separation pairs, sub-arcsecond resolution imaging at multiple wavelengths is necessary. Therefore, we obtained \hstname~(\hst) imaging for \offAGNhstsztext~galaxies identified using the method from \paperI. This paper is structured as follows: in Section \ref{sec:target_sample} we describe the \hst~targets and observations; in Section \ref{sec:nature_xray} we put constraints on the merger scenarios for each galaxy; in Section \ref{sec:selector} we discuss spatially offset X-ray sources as a selector of galaxy mergers hosting AGN; in Section \ref{sec:co-evol} we discuss offset AGN in the context of galaxy-SMBH co-evolution; and in Section \ref{sec:conclusions} we present our conclusions. Throughout we assume a cosmology defined by $H_{0}=70$ km s$^{-1}$ Mpc$^{-1}$, $\Omega_{M}=0.3$, and $\Omega_{\Lambda}=0.7$. \\

\begin{deluxetable}{ccccc}
\tabletypesize{\footnotesize}
\tablecolumns{5}
\tablecaption{Observing Details of the \hst~Targets.}
\tablehead{
\colhead{\TARGname \vspace*{0.05in}} &
\colhead{Redshift} &
\colhead{$t_{H}$} &
\colhead{$t_{I}$} &
\colhead{$t_{B}$} \\
\colhead{$-$ \vspace*{0.05in}} &
\colhead{$-$} &
\colhead{(\uSEC)} &
\colhead{(\uSEC)} &
\colhead{(\uSEC)} \\
\colhead{(1)} &
\colhead{(2)} &
\colhead{(3)} &
\colhead{(4)} &
\colhead{(5)}
}
\startdata
SDSS J081330.15$+$541844.4 & 0.041 & 147 & 918 & 1092 \\
SDSS J094032.25$+$311328.5 & 0.170 & 147 & 618 & 1329 \\
SDSS J102141.89$+$130550.4 & 0.077 & 147 & 618 & 1302 \\
SDSS J111458.01$+$403611.4 & 0.076 & 147 & 618 & 1329 \\
SDSS J123420.12$+$475155.7 &  0.183 & 147 & 918 & 1041 \\
SDSS J212512.48$-$071329.9 & 0.064 & 147 & 918 & 999
\enddata
\tablecomments{Column 1: Target galaxy name; Column 2: SDSS spectroscopic redshift; Column 3: \Hb-band image exposure time; Column 4: \Ib-band image exposure time; and Column 5: \Bb-band image exposure time.}
\label{tab:targets}
\end{deluxetable}

\section{The \hst~Target Sample}
\label{sec:target_sample}

The main sample analyzed in this paper consists of \offAGNhstsztext~galaxies identified using the method from \paperI~that were observed by \hst~(Program: GO 14068, PI: Barrows). Full details of the sample selection are provided in \paperI, but here we provide the basic properties: each galaxy was originally observed by the \sdss~Data Release 7 spectroscopic fiber survey and has optical emission line flux ratios of \oiii/\hb~and \niib/\ha~that place them in the AGN region of a typical Baldwin-Phillips-Terlevich (BPT) diagram \citep{Baldwin1981} as defined in \citet{Kewley:2006}. This diagnostic suggests that an AGN is present within the $1\farcs5$ fiber radius centered on the galaxy. Furthermore, an X-ray source with a rest-frame, unabsorbed $2-10$ keV luminosity (\LXhard) or hardness ratio consistent with the presence of an AGN is also detected and has a $1\sigma$ error ellipse that spatially overlaps with the SDSS fiber. We then selected galaxies where the X-ray source is spatially offset from the galaxy centroid based on SDSS imaging. Since AGN are expected to reside in the nuclei of galaxies, the spatial offsets mark these sources as galaxy merger candidates. Due to the selection requirement that the X-ray sources overlap with the $1\farcs5$ fiber radius, the resolution of the SDSS imaging (FWHM~$=1\farcs6$) is not sufficient to detect secondary stellar cores associated with the offset X-ray sources.

\begin{figure*}[t!] $
\begin{array}{c c c}
\hspace*{-0.0in} \includegraphics[width=0.32\textwidth]{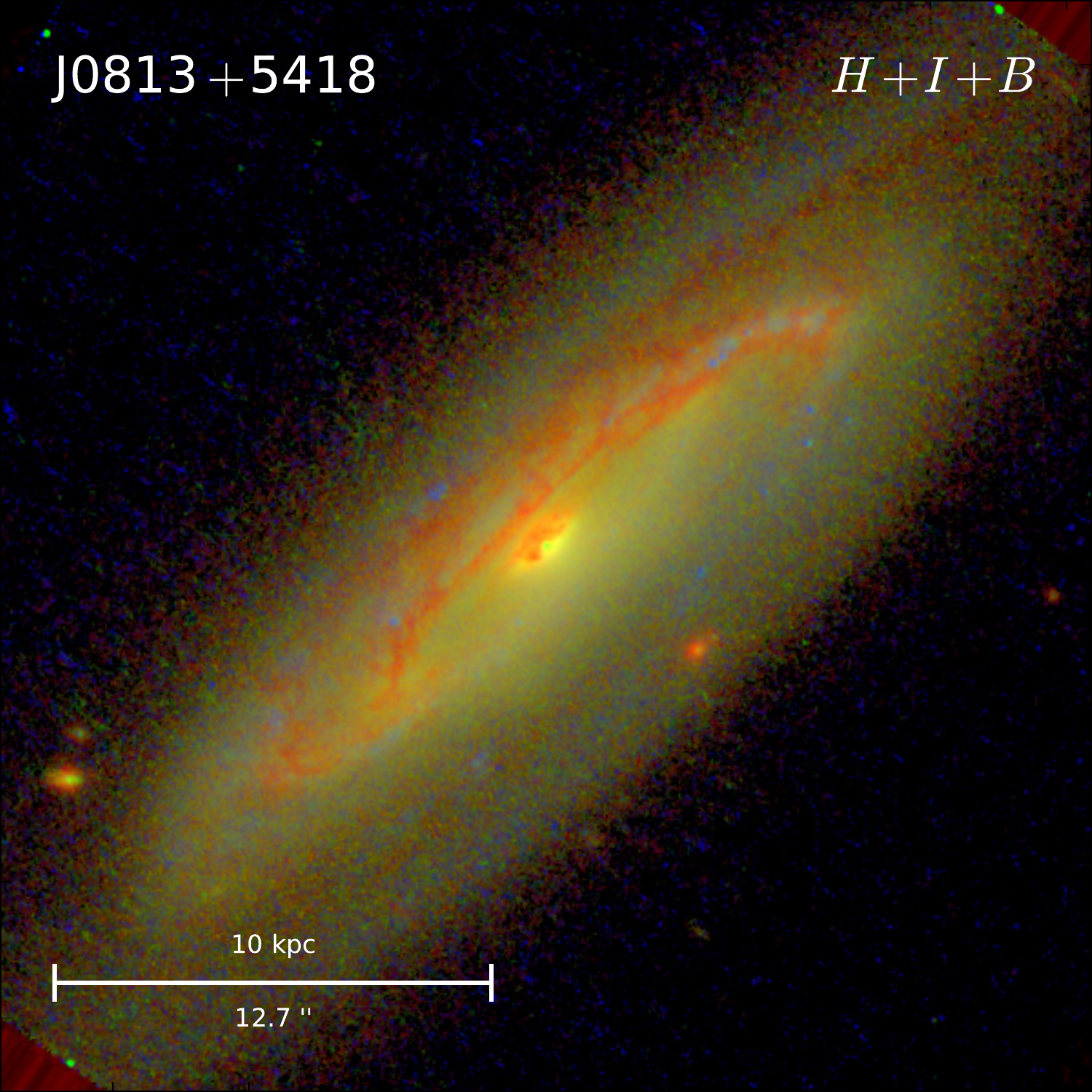} &
\hspace*{-0.0in} \includegraphics[width=0.32\textwidth]{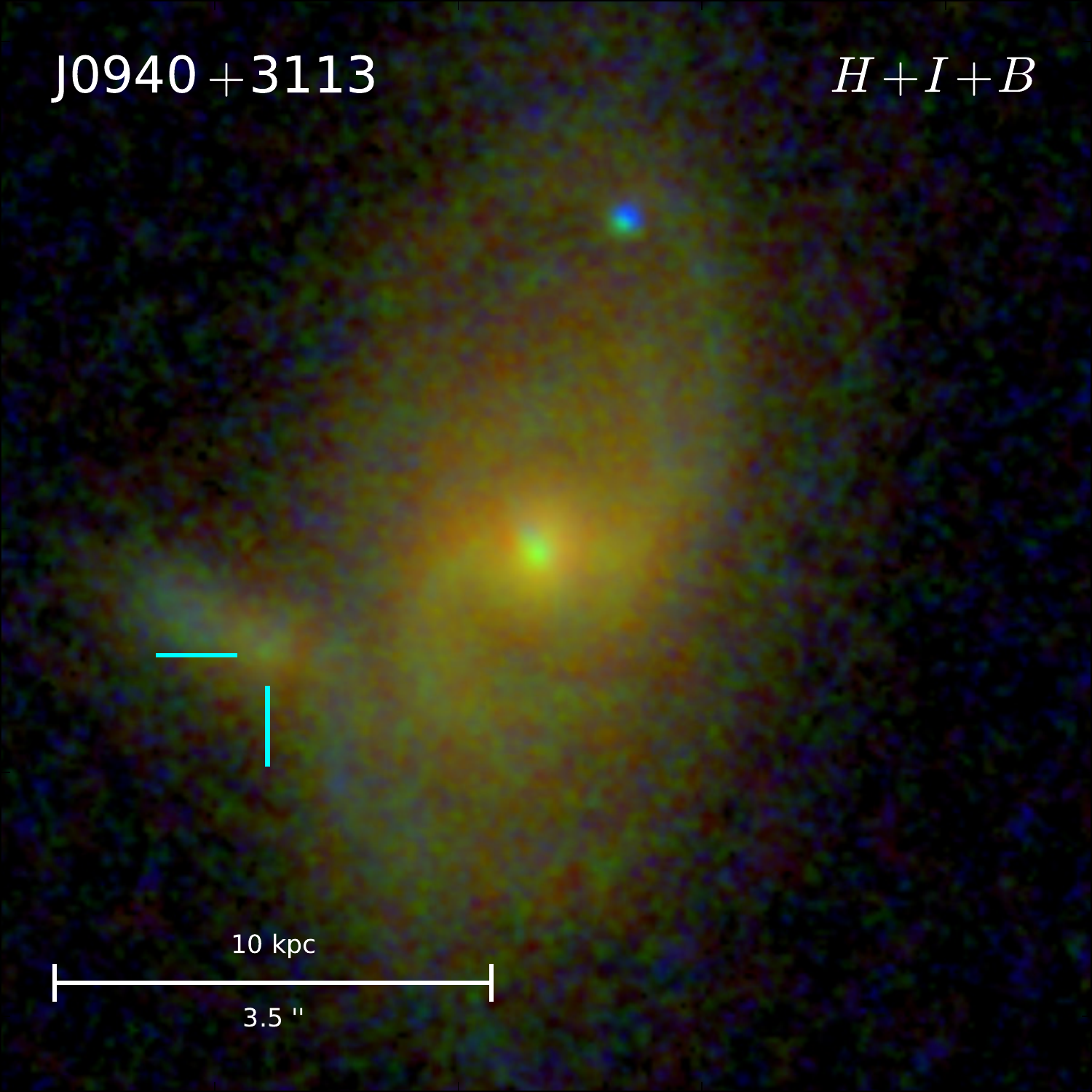} &
\hspace*{-0.0in} \includegraphics[width=0.32\textwidth]{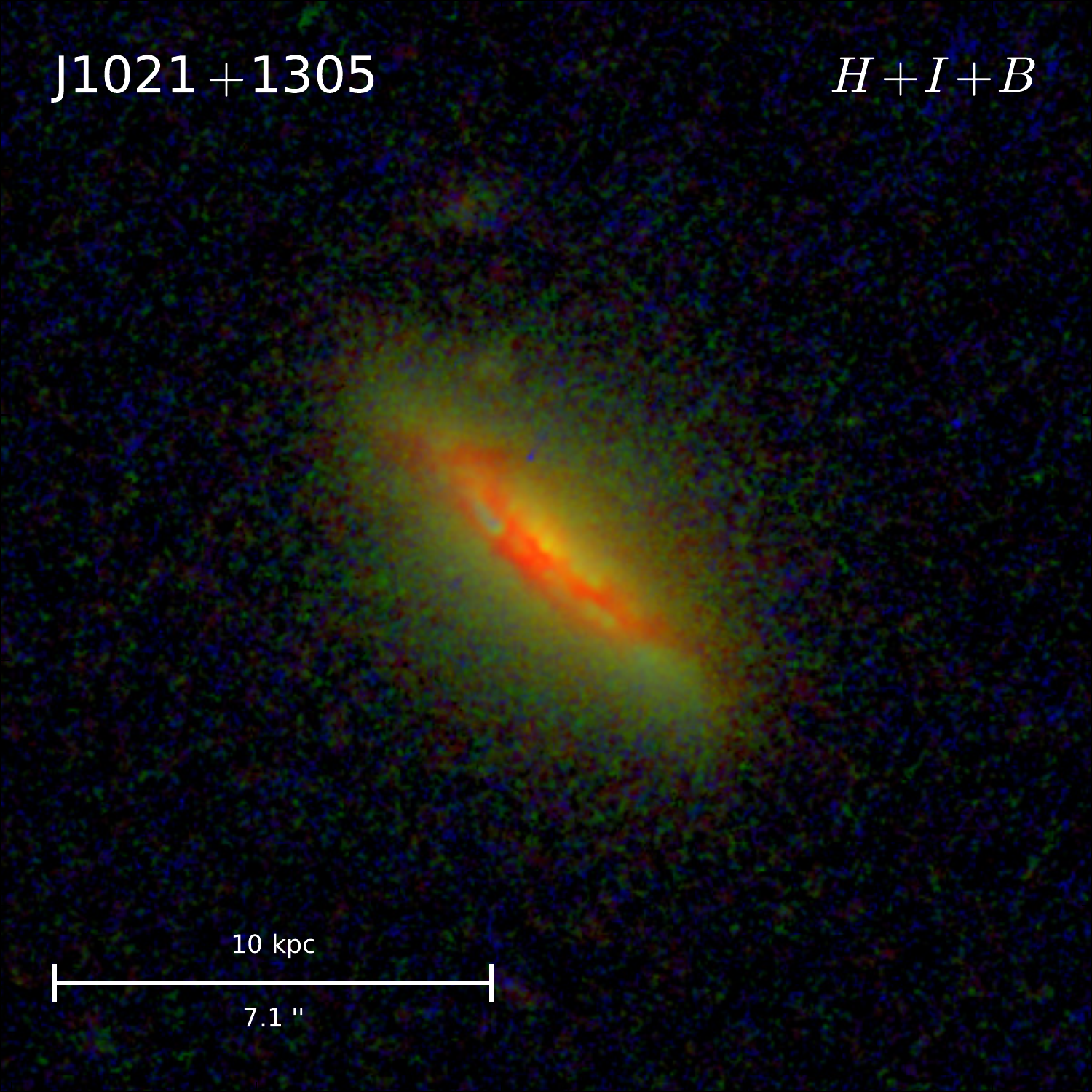} \\
\hspace*{-0.0in} \includegraphics[width=0.32\textwidth]{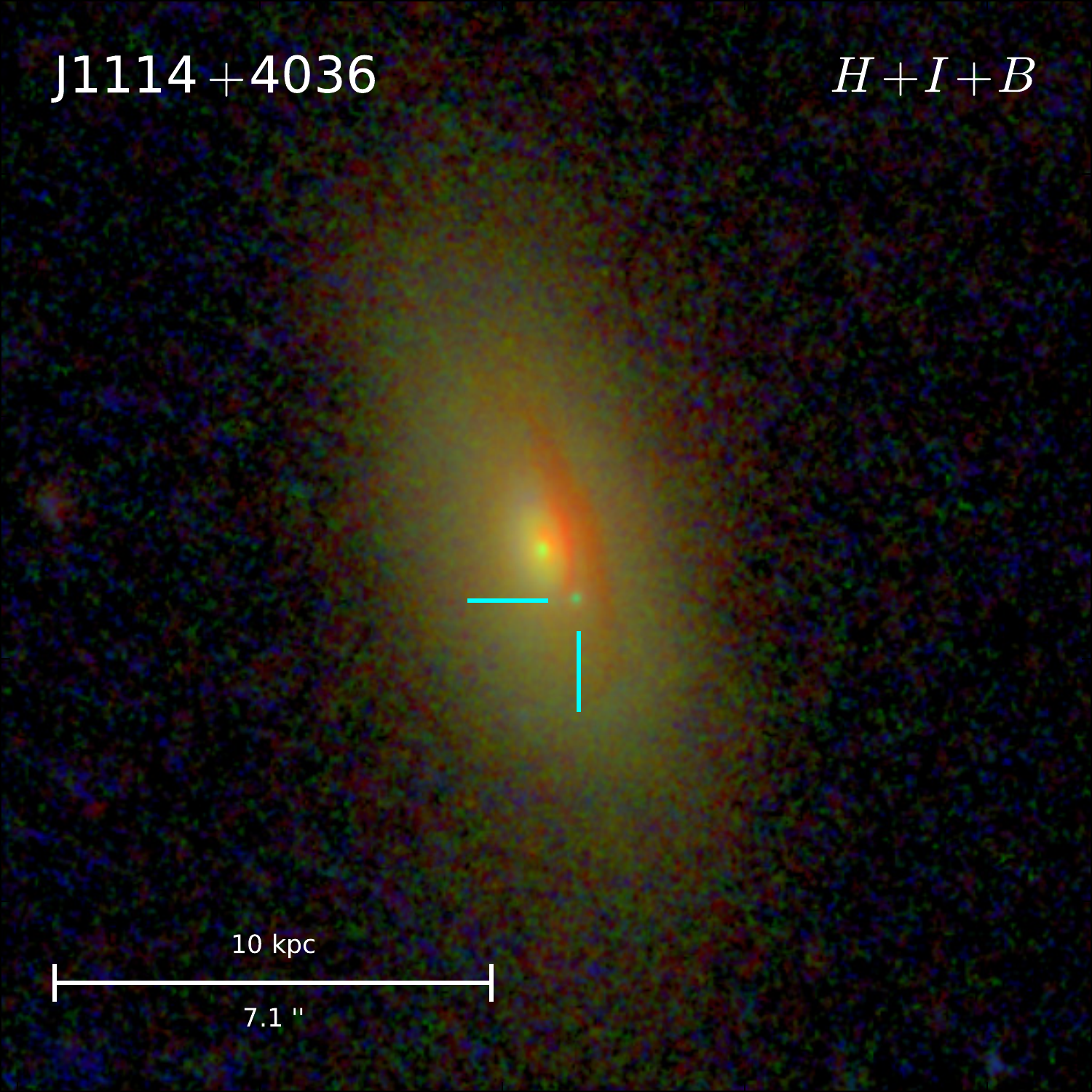} &
\hspace*{-0.0in} \includegraphics[width=0.32\textwidth]{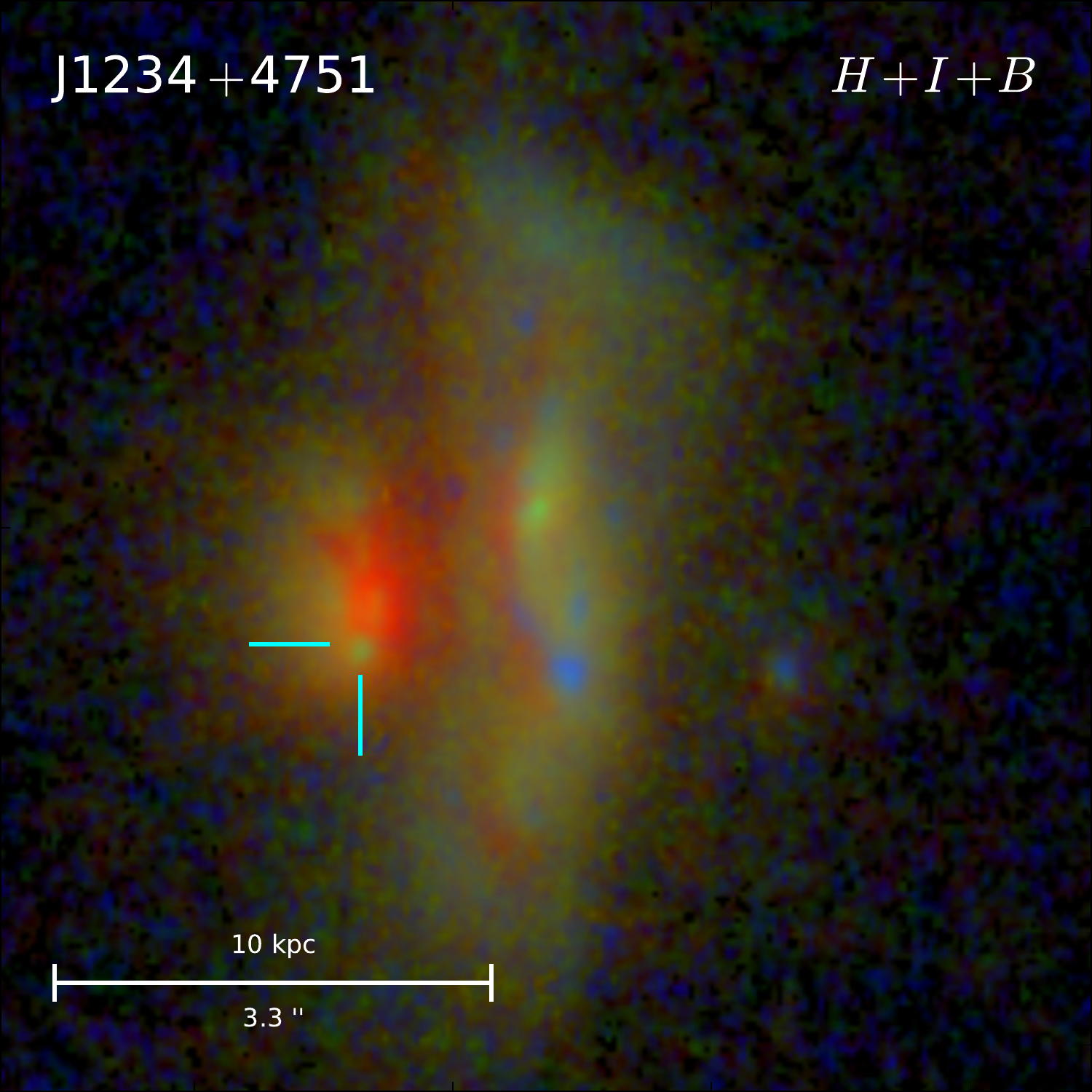} &
\hspace*{-0.0in} \includegraphics[width=0.32\textwidth]{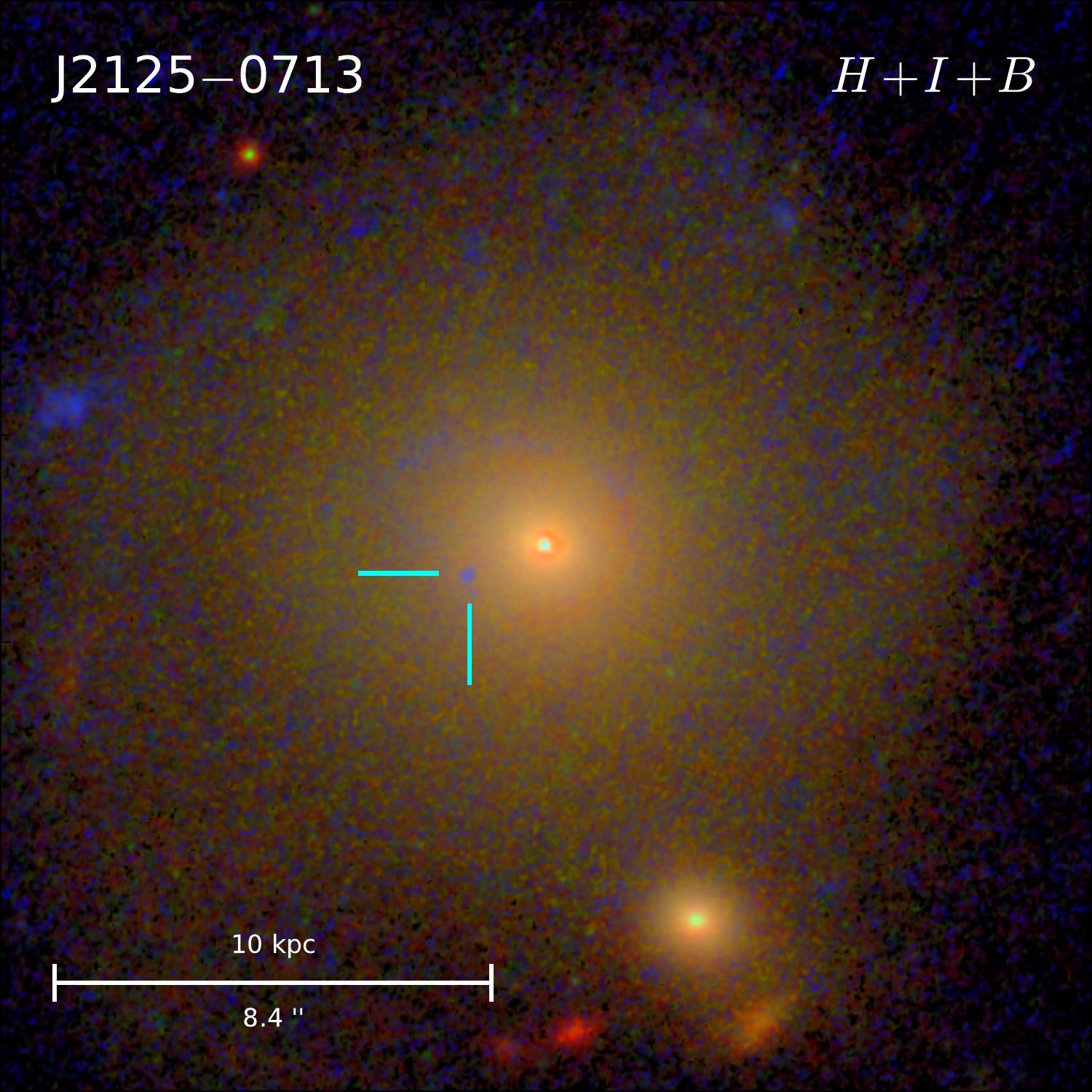}
\end{array} $
\caption{\footnotesize{$H+I+B$ color composite images of the \offAGNhstsztext~\hst~targets. The \Hb-, \Ib-, and \Bb-band counts are displayed on red, green, and blue colors scales, respectively. The FOV in each panel is \fitdim$\times$\fitdim~kpc (\gf~fitting box size; Section \ref{sec:detect_core}) with North up and East to the left. Note that four of the galaxies (\offtwo, \offfour, \offfive, and \offsix) show evidence for interacting \second~stellar cores as described in Section \ref{sec:mergers} (marked with cyan crosshairs) while the remaining two (\offone~and \offthree) do not.
}
}
\label{fig:HST_RGB}
\end{figure*}

Each target was imaged with three \hst/\wfcthree~wide band filters: \Hb~(F160W; \lampeak~$=$~\Hlampeak), \Ib~(F814W; \lampeak~$=$~\Ilampeak), and \Bb~(F438W; \lampeak~$=$~\Blampeak). The \Hb-band filter was chosen to provide a continuum map of the older stars typically found in galaxy bulges hosting SMBHs. The \Ib- and \Bb-band filters were chosen to provide continuum maps of younger stars and to trace optical emission lines and scattered light from star formation or AGN. The target names, SDSS spectroscopic redshifts, and observation details (filter and exposure time) are listed in Table \ref{tab:targets}. The targets were chosen because they had not previously been observed by \hst.

\section{Galaxy Mergers Revealed by \hst}
\label{sec:nature_xray}

While our selection of spatially offset X-ray sources can reach down to physical separations of $\sim1$ kpc, these sizes correspond to angular scales of $<1''$ in some cases. Hence, \hst~imaging is vital for understanding if the offset X-ray sources are related to galaxy mergers. Figure \ref{fig:HST_RGB} shows the $H+I+B$ color composite images of the \offAGNhstsztext~\hst~targets. While the SDSS images used in the original selection provide no explicit evidence of mergers, the \hst~imaging reveals signatures of on-going galaxy mergers in the form of \second~stellar cores that are spatially offset from the \prim~(i.e. more massive) galaxy in four of the targets (\offtwo, \offfour, \offfive, and \offsix). We emphasize that each \second~stellar core was previously undetected in the SDSS imaging. The other two galaxies (\offone~and \offthree) do not show evidence of \second~stellar cores or merger related morphological disturbances. Thus, the \hst~imaging reveals that $4/6$ systems are on-going galaxy mergers. In two of those systems (\offfour~and \offfive) an offset X-ray source is spatially coincident with the \second~stellar core and hence the merger selection is a direct result of the spatially offset X-ray source detection. In the other two systems (\offtwo~and \offsix) the X-ray source is most likely associated with the \prim~stellar core detected in the \hst~imaging. The implications of these results are discussed in Section \ref{sec:selector}.

\subsection{Detecting Stellar Cores}
\label{sec:detect_core}

We use the \Hb-band images to detect stellar cores associated with the primary galaxy and potential companion galaxies in each of the \offAGNhstsztext~targets. The \Hb-band images are dominated by light from near-infrared (NIR) stellar continuum emission that corresponds to relatively old stars in galaxy bulges \citep{Mannucci:2001}. Therefore, we fit each system with a combination of multiple two-dimensional Sersic functions \citep{Sersic:1968} that have been empirically demonstrated to be reliable descriptions of galaxy stellar bulges \citep{Graham:2005}. Moreover, in \citet{comerford:2015} we showed that the Sersic centroid fit to the individual stellar cores of a galaxy merger is a robust tracer of the peak \Hb-band brightness, regardless of the residuals at large radii from the centroid.

The full procedure is described in \citet{Barrows:2017b} though here we provide a basic description. First, we run \se~\citep{Bertin:Arnouts:1996} on the \Hb-band images to generate a base list of all the significantly detected ($>3\sigma$) sources. From this list, we use \gf~\citep[version 3.0.5;][]{Peng:2010} to fit Sersic components to all of the detected sources within a \fitdim$\times$\fitdim~kpc field-of-view (FOV) centered on the SDSS J2000 right ascension (RA) and declination (DEC) of the galaxy, plus a uniform sky component. The FOV is chosen to allow all nearby contaminating sources to be included in the \Hb-band fitting box and modeled for all \offAGNhstsztext~galaxies. Since the \sdss~fiber spectra classify each system as a \typeII~AGN, the NIR AGN contribution is expected to be weak. Indeed, we find that PSF components are not statistically warranted in any cases. However, in some cases Sersic components did not converge on the \se~detections. In these instances we manually fit the sources with two-dimensional Gaussian functions.

\subsection{Four Galaxy Mergers}
\label{sec:mergers}

Figure \ref{fig:HST_H_fiber} shows the \Hb-band images zoomed in on the SDSS fiber position. The brightest model component is considered to be the \prim~stellar core, and in all cases it is the component nearest the centroid of the target galaxy from SDSS imaging. To reveal fainter stellar core detections, Figure \ref{fig:HST_H_fiber} also shows the residuals after subtracting the model component associated with the \prim~galaxy from each image. In four of the \offAGNhstsztext~systems (\offtwo, \offfour, \offfive, and \offsix) the modeling procedure detects a \second~stellar core that is $<10$ kpc from the \prim~galaxy and physically interacting with it based on proximity and/or visually apparent connecting stellar features.

The locations of the \prim~and \second~stellar cores are marked in Figure \ref{fig:HST_H_fiber}. The physical separations of the stellar cores (\deltas) have a range of \deltas~$=$~\fourphysep$-$\twophysep~kpc and are listed in Table \ref{tab:gal_props}. In the following sections we use the \hst~imaging to estimate stellar masses (Section \ref{sec:major_minor}), determine whether the X-ray sources are in the more or less massive galaxy (Section \ref{sec:loc_agn}), and examine the colors for evidence of nuclear obscuration or additional AGN that are undetected in X-rays (Section \ref{sec:dual_agn}).

\begin{figure*}[t!] $
\begin{array}{c}
\hspace*{0.in} \includegraphics[width=\textwidth]{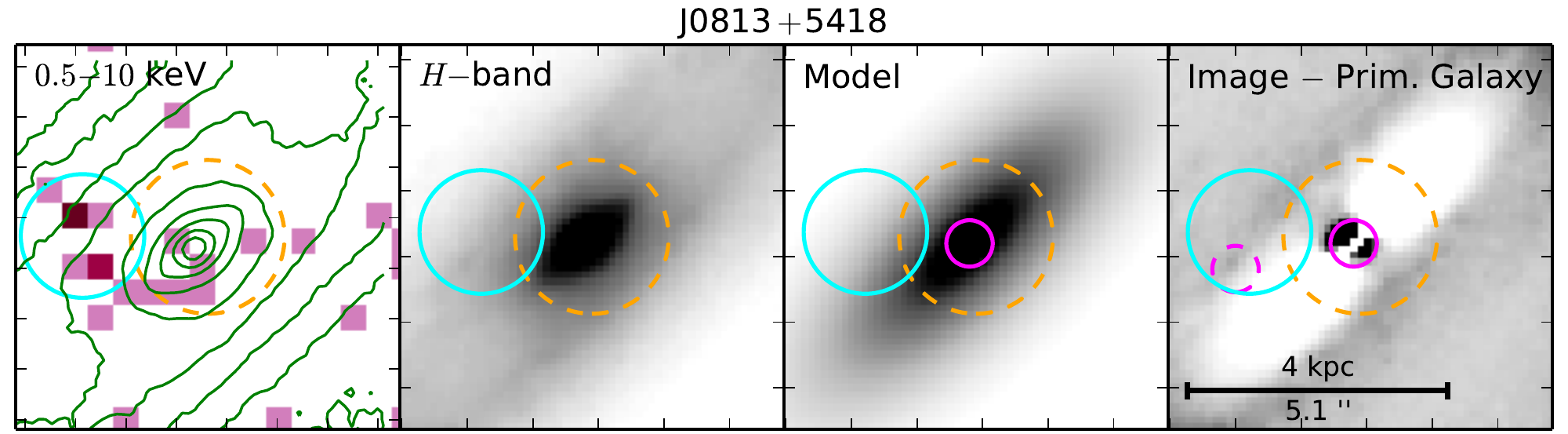} \\
\hspace*{-0.in} \includegraphics[width=\textwidth]{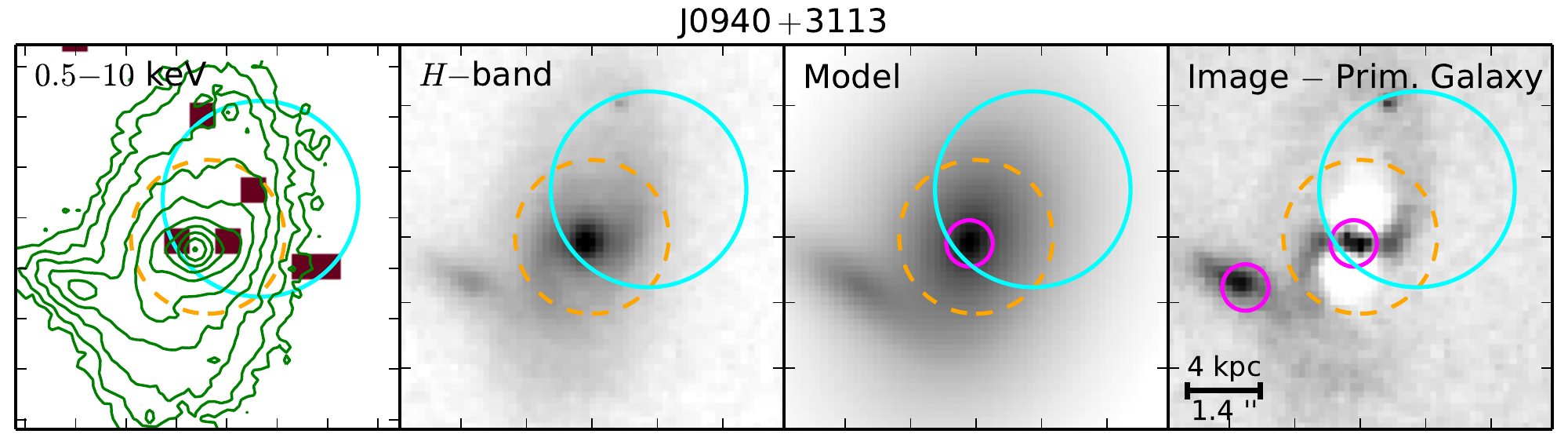} \\
\hspace*{0.in} \includegraphics[width=\textwidth]{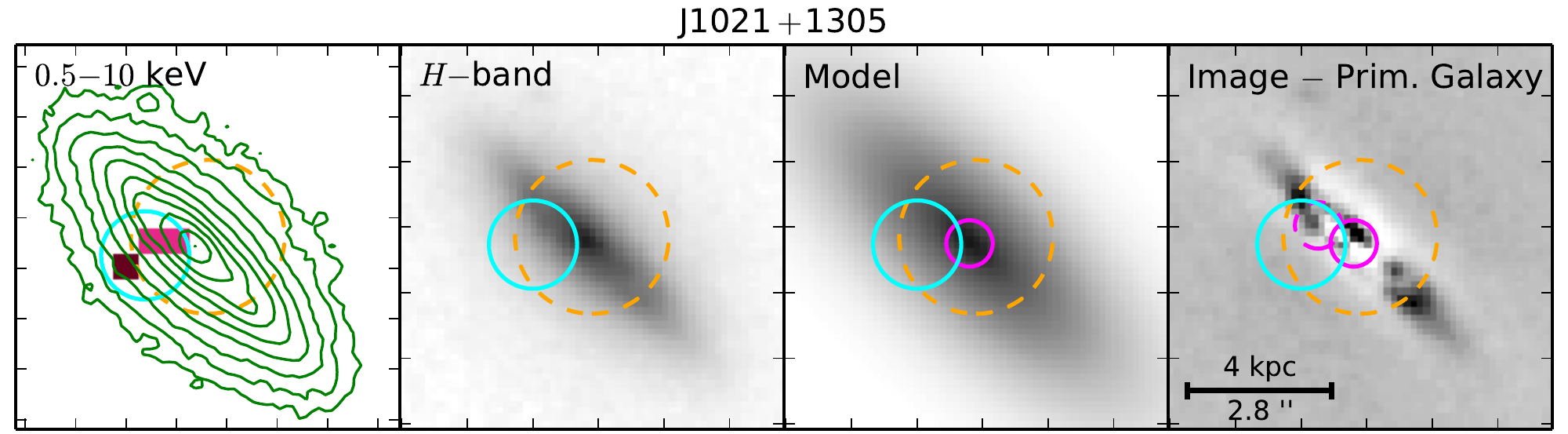}
\end{array}$
\caption{\footnotesize{Images of the \offAGNhstsztext~\hst~targets focused on the SDSS fiber (orange, dashed circles) with North up and East to the left. From left to right, the panels are as follows: rest-frame $0.5-10$ keV \ch~image (un-binned) and \Hb-band image contours in green, \Hb-band image, \Hb-band \gf~model, and the residuals obtained by subtracting the \prim~model component from the image. The solid magenta circles denote the \prim~stellar core (fiber center) and the \second~stellar core. The dashed magenta circles denote upper limits on stellar core detections within $1\sigma$ of the X-ray source position. The magenta circle sizes represent the apertures used for extraction of fluxes in Section \ref{sec:dual_agn}. The cyan circles represent the X-ray source position and $1\sigma$ uncertainties (combined uncertainties of the model centroids and the relative astrometry).
}
}
\label{fig:HST_H_fiber}
\end{figure*}

\begin{figure*}[t!] $
\ContinuedFloat
\begin{array}{c}
\hspace*{-0.in} \includegraphics[width=\textwidth]{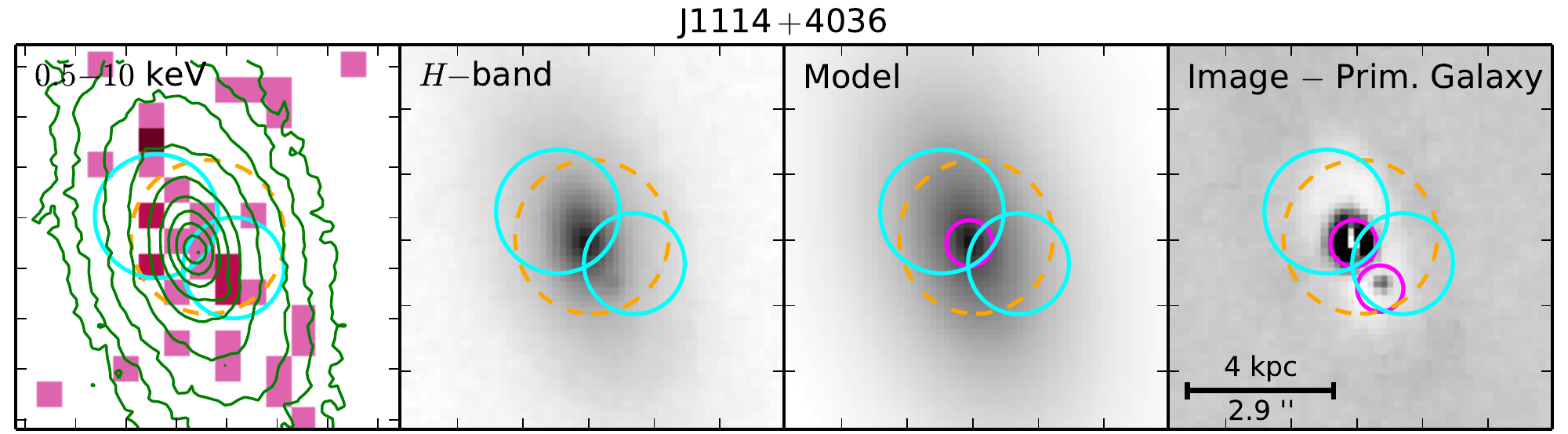} \\
\hspace*{0.in} \includegraphics[width=\textwidth]{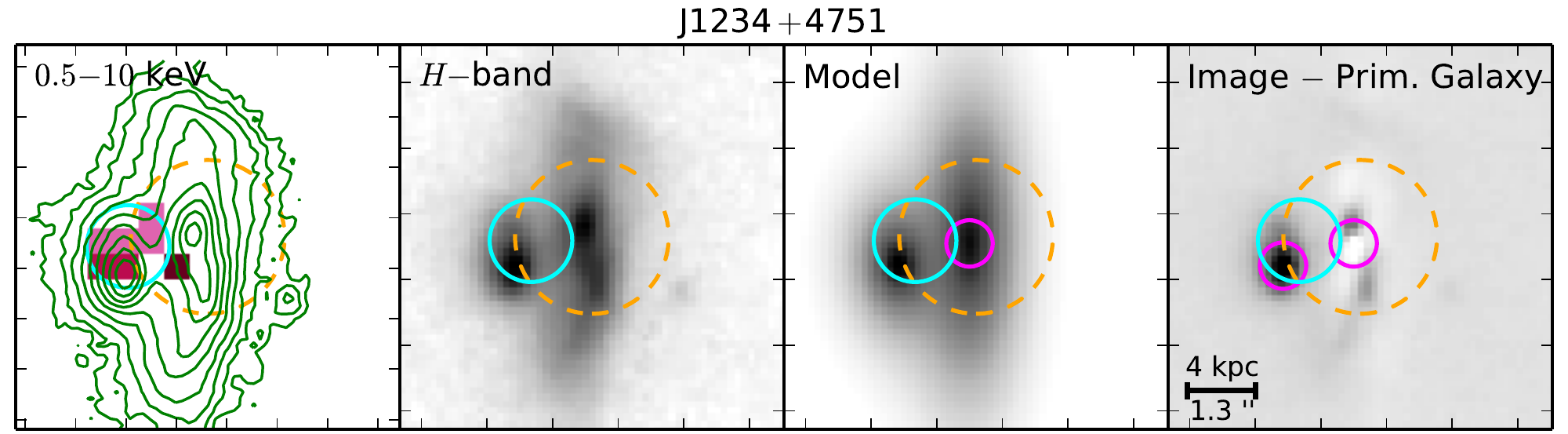} \\
\hspace*{-0.in} \includegraphics[width=\textwidth]{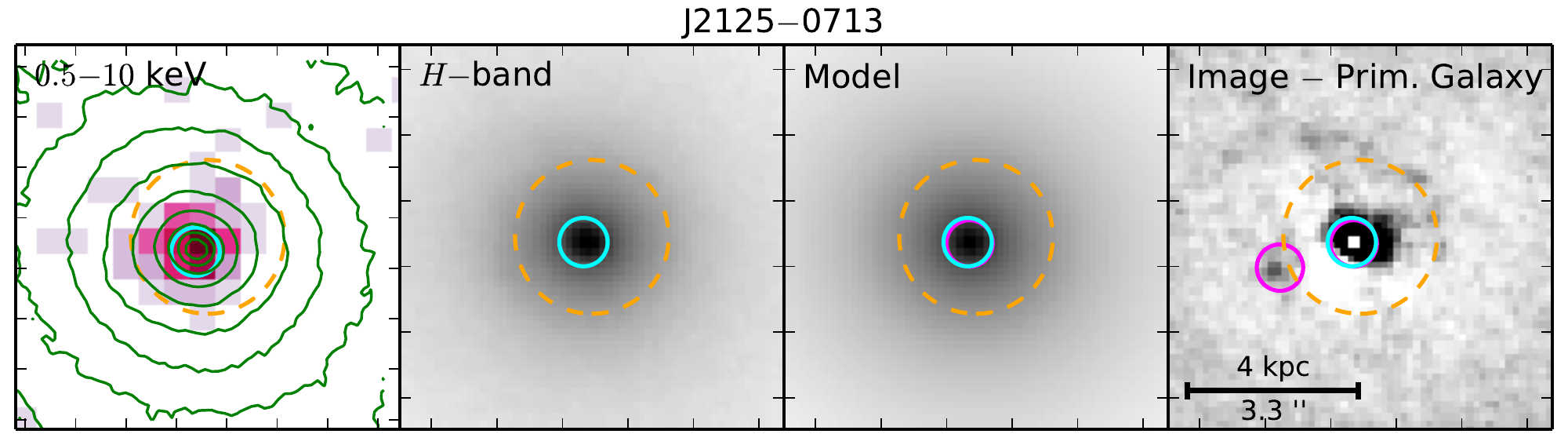}
\end{array}$
\caption{\footnotesize{continued.
}
}
\end{figure*}

\subsubsection{One Major Merger and Three Minor Mergers}
\label{sec:major_minor}

Since the \Hb-band fluxes of the \prim~and \second~stellar cores trace continuum emission from bulge stars, they are assumed to be proportional to the stellar masses (\Mone~and \Mtwo, respectively). Therefore, we use the \Hb-band flux ratios between the \prim~and \second~stellar cores as proxies for the mass ratios (\mratio). The mass ratios have a range of \mratio~$=$~\fivemratio$-$\sixmratio~and are listed in Table \ref{tab:gal_props}. When adopting a mass ratio of \MRATIOthresh~as the division between major mergers (\mratio~$<$~\MRATIOthresh) and minor mergers (\mratio~$>$~\MRATIOthresh), \offfive~is classified as a major merger while \offtwo, \offfour, and \offsix~are classified as minor mergers.

The values of \Mone~and \Mtwo~are estimated from the mass ratios and the total galaxy stellar masses (\Mgal). \Mgal~is measured from the mass-to-light ratio function of \citet{Bell:2003} based on the SDSS \sdssg-\sdssr~colors, the flux from the SDSS \sdssz~filter (SDSS filter that most closely traces the continua of older stars that reside in galaxy bulges), and the \sdssz~filter $k-$correction. The values of \Mone~and \Mtwo~are listed in Table \ref{tab:gal_props}. The \second~stellar cores have masses that range from \Mtwo~$=$~\sixHtwoMgal~$-$~\fiveHtwoMgal~\Msun. Assuming that the \second~stellar cores are the remnant bulges of galaxies, the estimated black hole (BH) masses (\MBH) range from \MBH~$=$~\sixHtwoMBH~$-$~\fiveHtwoMBH~\Msun~using the empirical relation between BH mass and bulge mass from \citet{Marconi:Hunt:2003}. These BH mass estimates are within the range of values typically measured for the SMBHs of galaxy nuclei (\MBH~$=10^{5}-10^{9}$ \Msun).

Since the galaxy mass estimates are based on the image modeling components, they do not account for deviations from Sersic or Gaussian profiles at large radii or for mass exchange during the merger. In particular, the host galaxies of the \second~stellar cores may have experienced significant tidal stripping. Therefore, the original masses of the \second~galaxies were likely higher.

\subsubsection{Locations of the X-ray Sources}
\label{sec:loc_agn}

We register the \Hb-band images with the \ch~images of the X-ray sources (Figure \ref{fig:HST_H_fiber}) to determine where the AGN are located within the merging systems. The registration procedure is based on \paperI~and further described with respect to \hst~imaging in \citet{Comerford:2017a,Comerford:2017b}. The X-ray source positions are measured using the image modeling procedure described in \paperI~and are marked in Figure \ref{fig:HST_H_fiber}. The X-ray sources are detected at $>3\sigma$ significance in four of the galaxies (\offone, \offfour, \offfive, and \offsix) and $>2\sigma$ significance in the remaining two (\offtwo~and \offthree). We consider X-ray sources to be associated with stellar cores if they are spatially coincident within $1\sigma$ when accounting for the centroid errors and relative astrometric uncertainties. Three of the galaxy mergers have only one X-ray source (offset AGN), and in two of those systems (\offtwo~and \offsix) the X-ray source is associated with the \prim~galaxy, while in the other system (\offfive) the X-ray source is associated with the \second~stellar core.

The \ch~image of the fourth system (\offfour) is best fit by a model that consists of two X-ray sources. Both sources are consistent with accretion onto massive BHs based on their rest-frame hard X-ray luminosities (\LXhard~$>10^{41}$ \uLum). While the centroid errors of each source are relatively large, one of the X-ray sources is statistically consistent with the \prim~galaxy nucleus while the other is consistent with the \second~stellar core based on the $1\sigma$ uncertainties. When considering that two \LXhard~$>10^{41}$ \uLum~X-ray sources are likely present, and that they are each spatially coincident with a stellar core, we consider this system to be a dual AGN candidate. The X-ray source positions, source counts \citep[computed using the Bayesian Estimation of Hardness Ratios procedure;][]{Park:2006}, model detection significances, and \LXhard~values are listed in Table \ref{tab:xray_props} for all \offAGNhstsztext~\hst~targets. The X-ray source spatial offsets from each stellar core (including the offset position angles) are listed in Table \ref{tab:xray_offsets} for all \offAGNhstsztext~\hst~targets.

\begin{deluxetable*}{lcccccccccc}
\tabletypesize{\footnotesize}
\tablecolumns{11}
\tablecaption{Host Galaxy Properties.}
\tablehead{
\colhead{\name \vspace*{0.05in}} &
\colhead{\deltas} &
\colhead{\mratio} &
\colhead{\Mone} &
\colhead{\Mtwo} &
\colhead{\BIonecolor} &
\colhead{\BItwocolor} &
\colhead{\IHonecolor} &
\colhead{\IHtwocolor} &
\colhead{\BHonecolor} &
\colhead{\BHtwocolor} \\
\colhead{$-$ \vspace*{0.05in}} &
\colhead{(\uKPC)} &
\colhead{$-$} &
\colhead{(log[\Mone/\Msun])} &
\colhead{(log[\Mtwo/\Msun])} &
\colhead{$-$} &
\colhead{$-$} &
\colhead{$-$} &
\colhead{$-$} &
\colhead{$-$} &
\colhead{$-$} \\
\colhead{(1)} &
\colhead{(2)} &
\colhead{(3)} &
\colhead{(4)} &
\colhead{(5)} &
\colhead{(6)} &
\colhead{(7)} &
\colhead{(8)} &
\colhead{(9)} &
\colhead{(10)} &
\colhead{(11)}
}
\startdata
J0813$+$5418 & $-$ & $-$ & $11.1$ & $-$ & $2.6$ & $2.1$ & $1.0$ & $1.2$ & $3.6$ & $3.3$\\
J0940$+$3113 & $6.6$ & $15$ & $10.8$ & $9.6$ & $2.5$ & $2.0$ & $0.8$ & $0.8$ & $3.3$ & $2.8$\\
J1021$+$1305 & $-$ & $-$ & $10.7$ & $-$ & $2.9$ & $2.3$ & $1.6$ & $1.6$ & $4.6$ & $3.9$\\
J1114$+$4036 & $1.2$ & $48$ & $10.7$ & $9.0$ & $2.1$ & $1.9$ & $0.9$ & $1.0$ & $3.0$ & $2.8$\\
J1234$+$4751 & $4.3$ & $2.0$ & $11.0$ & $10.7$ & $2.0$ & $2.8$ & $0.9$ & $1.6$ & $2.9$ & $4.5$\\
J2125$-$0713 & $1.6$ & $1000$ & $11.1$ & $8.1$ & $2.1$ & $1.9$ & $0.9$ & $0.9$ & $3.1$ & $2.8$
\enddata
\tablecomments{Column 1: Galaxy name; Column 2: projected nuclear physical separation between the \prim~galaxy and \second~stellar core; Column 3: mass ratio of the \prim~galaxy to the \second~stellar core; Column 4: mass of the \prim~galaxy; Column 5: mass of the \second~stellar core; Columns 6-7: \prim~and \second~\BIcolor~colors; Columns 8-9: \prim~and \second~\IHcolor~colors; and Columns 10-11: \prim~and \second~\BHcolor~colors.}
\label{tab:gal_props}
\end{deluxetable*}

\subsubsection{Nuclear Obscuration}
\label{sec:dual_agn}

Gas and dust in the nuclei of galaxies can simultaneously fuel AGN while suppressing their optical and X-ray signatures through attenuation. Therefore, we use the three \hst~filters to examine the photometry of each stellar core in search of excessive red colors that may indicate nuclear dust or excessive blue colors that may originate from AGN emission. 

As shown in \citet{Barrows:2017b}, \Ib- and \Bb-band contributions from AGN continuum emission scattered off of gas or AGN-photoionized emission line gas is not negligible. The dynamics of on-going or past mergers likely subjects the gas to asymmetric kinematic motion. Therefore, we do not model the \Ib- and \Bb-band images with \gf~since the model components are empirically derived to model stars in virialized motion or well-ordered rotation. Instead, to measure the \Ib- and \Bb-band magnitudes of each stellar core we extract fluxes from apertures centered on the RA and DEC of the \prim~and \second~stellar cores as determined in Section \ref{sec:detect_core} after first registering them with the \Hb-band images. The aperture radii are designed to be \apsizefactext~times the \Hb-band image FWHM (\HbFWHM). Aperture fluxes are then summed after subtracting an average local background measured from directly adjacent annuli of width equal to one times the \Hb-band image FWHM. Aperture detections of \Ib- and \Bb-band counterparts to the stellar cores are present at significances of $>3\sigma$ in all four mergers. We also extract fluxes from apertures with offsets chosen randomly from a uniform distribution of radii within two times the stellar core spatial offsets but excluding the \prim~and \second~aperture positions. In all four mergers each random aperture detection is $<2\sigma$ in significance, further suggesting that the \Ib- and \Bb-band detections are associated with the stellar cores rather than unrelated features associated with hot gas or star formation. 

\begin{figure*} $
\begin{array}{cc}
\hspace*{0.in} \includegraphics[width=0.5\textwidth]{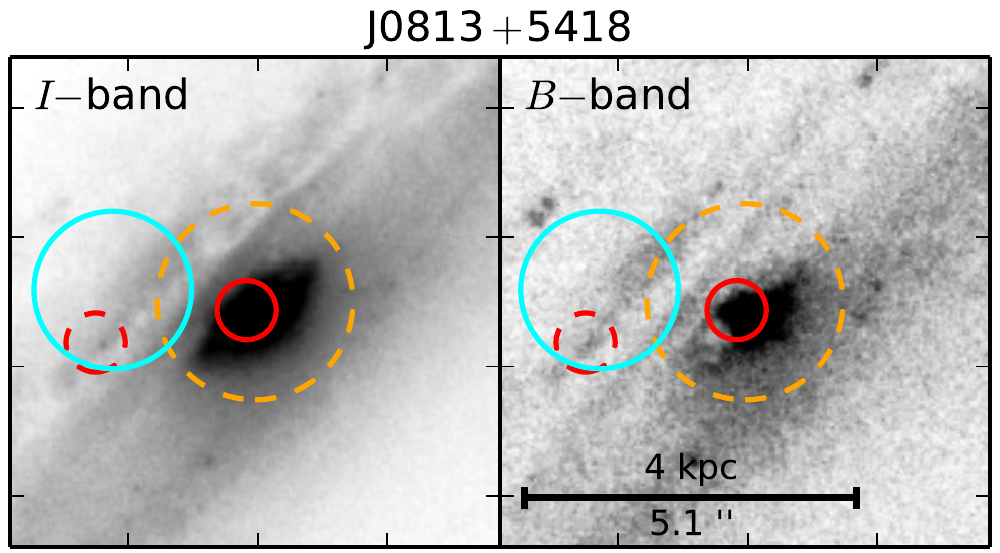} &
\hspace*{-0.1in} \includegraphics[width=0.5\textwidth]{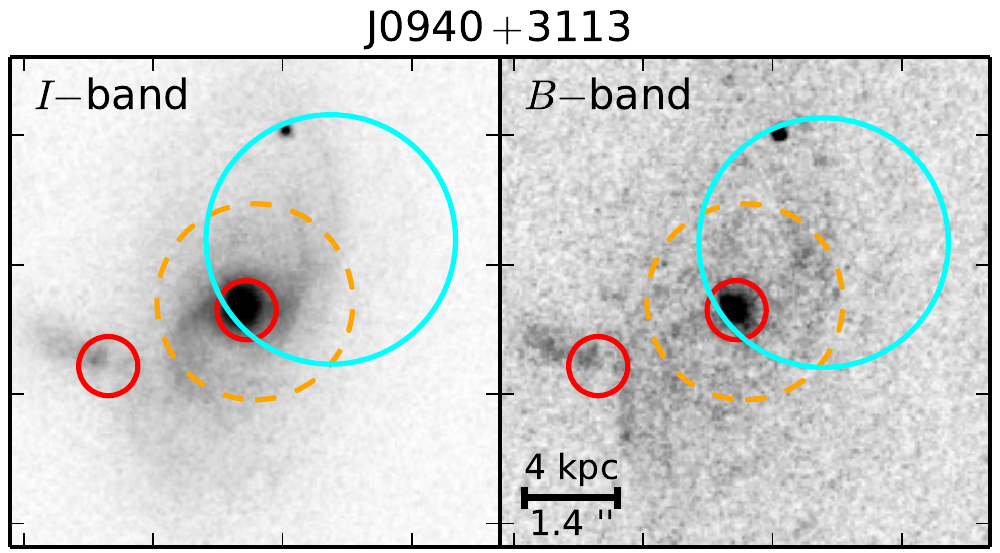} \\
\hspace*{0.in} \includegraphics[width=0.5\textwidth]{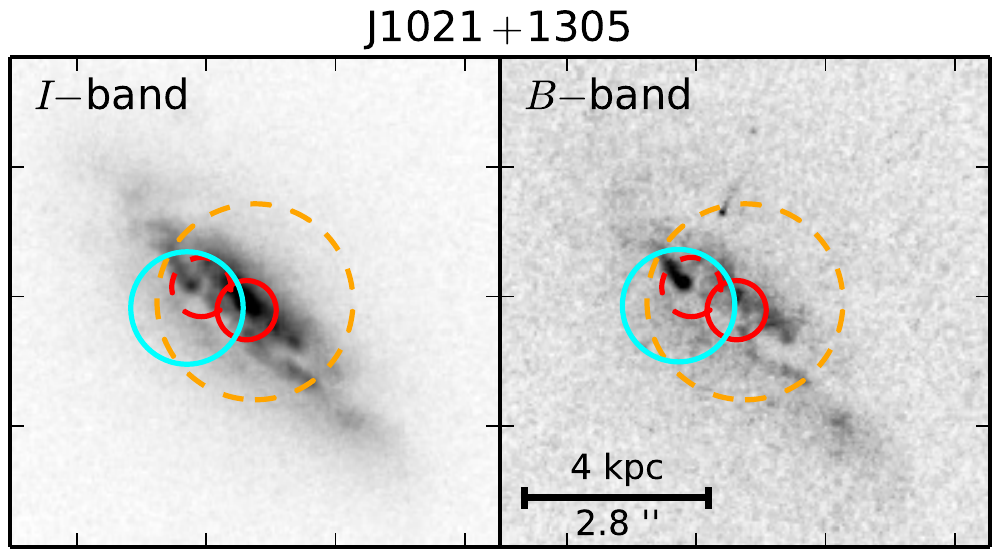} &
\hspace*{-0.1in} \includegraphics[width=0.5\textwidth]{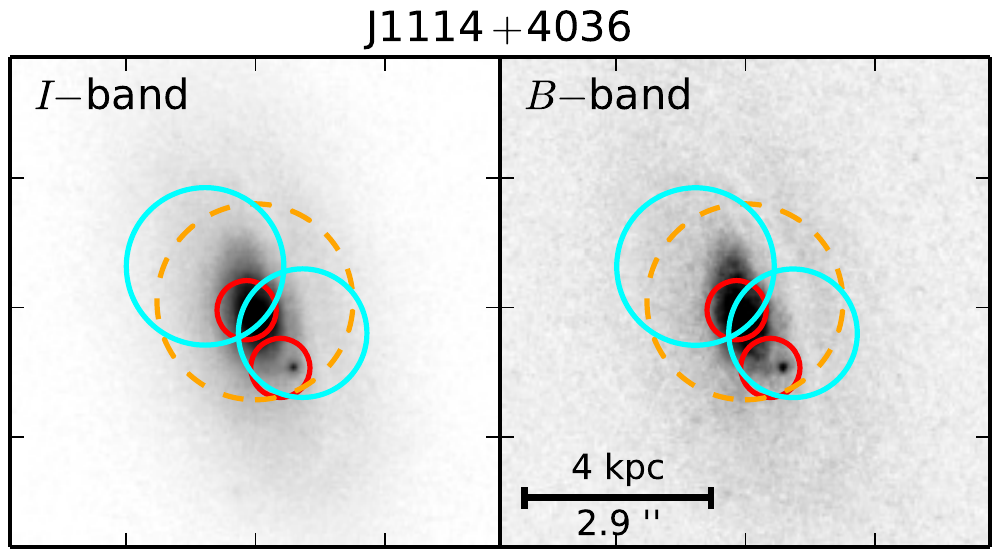}  \\
\hspace*{0.in} \includegraphics[width=0.5\textwidth]{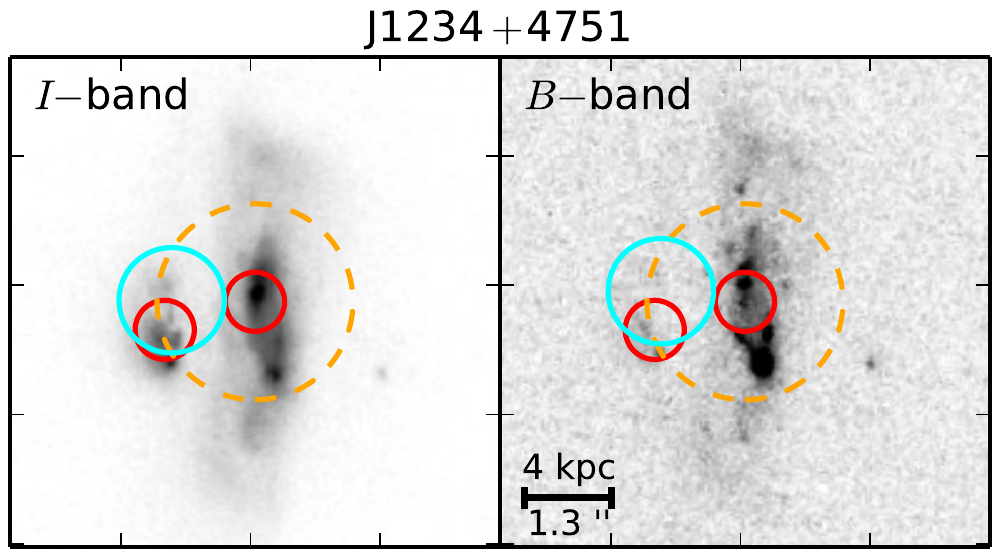} &
\hspace*{-0.1in} \includegraphics[width=0.5\textwidth]{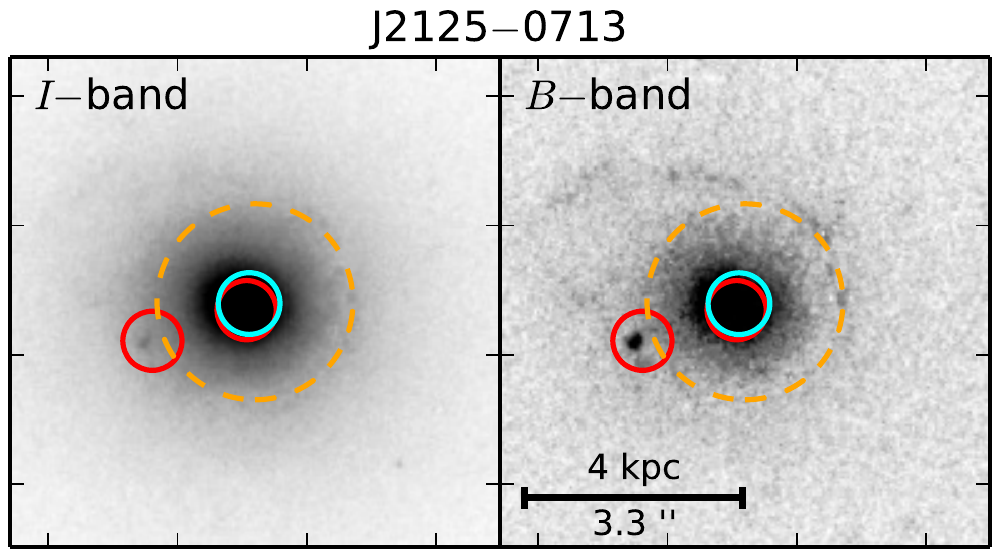}
\end{array}$
\caption{\footnotesize{\Ib-band images (left) and \Bb-band images (right) of the \offAGNhstsztext~\hst~targets focused on the SDSS fiber (orange, dashed circles) and displayed over the same FOV as in Figure \ref{fig:HST_H_fiber} with North up and East to the left. The solid magenta circles denote the \prim~stellar core (fiber center) and the \second~stellar core. The dashed magenta circles denote upper limits on stellar core detections within $1\sigma$ of the X-ray source position. The magenta circle sizes represent the apertures used for extraction of fluxes in Section \ref{sec:dual_agn}. The cyan circles represent the X-ray source position and $1\sigma$ uncertainties (combined uncertainties of the model centroids and the relative astrometry).}
}
\label{fig:HST_IB_fiber}
\end{figure*}

\begin{figure}[t!] 
\hspace*{-0.0in} \includegraphics[width=0.46\textwidth]{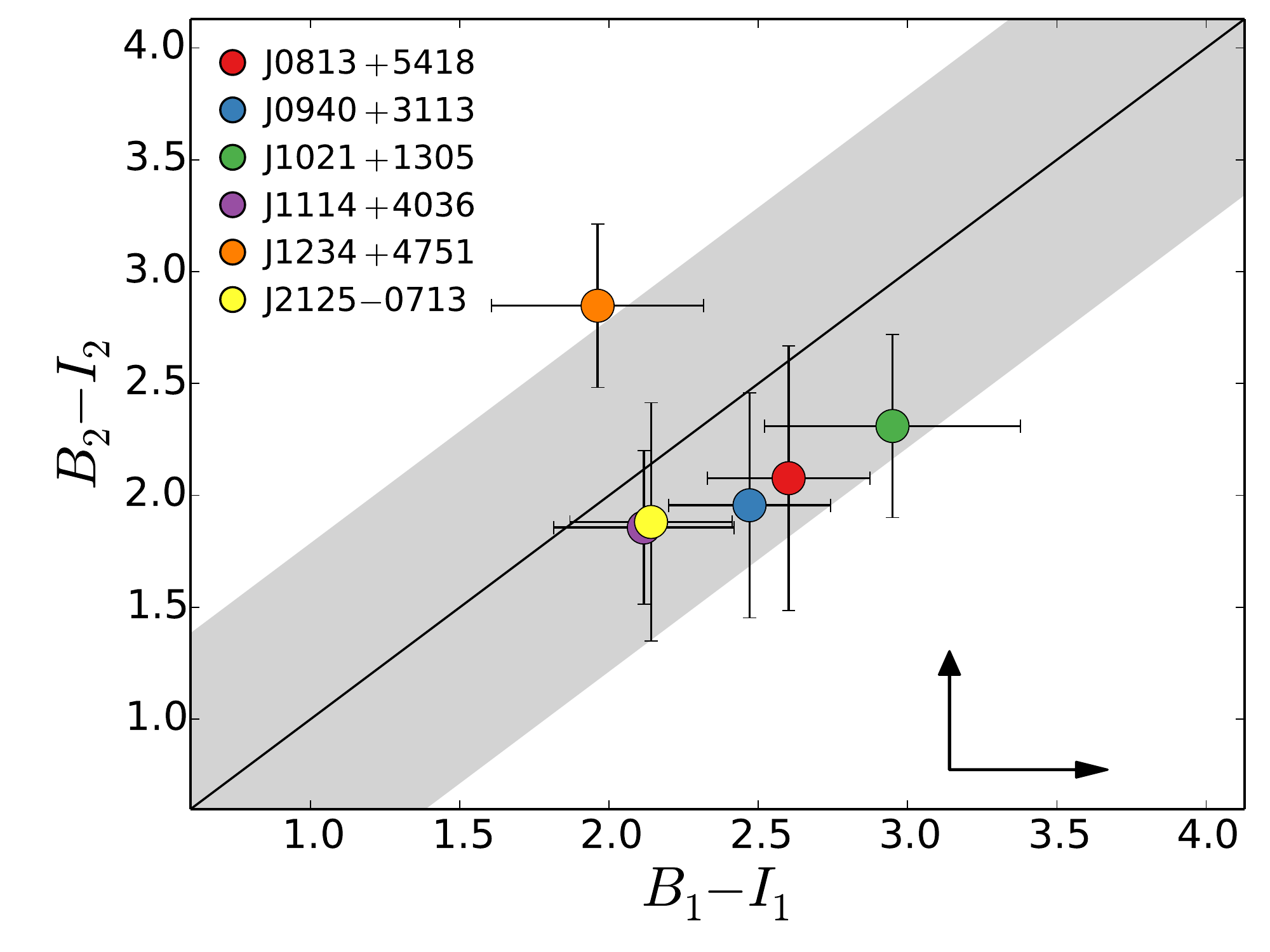} \\
\hspace*{-0.in} \includegraphics[width=0.46\textwidth]{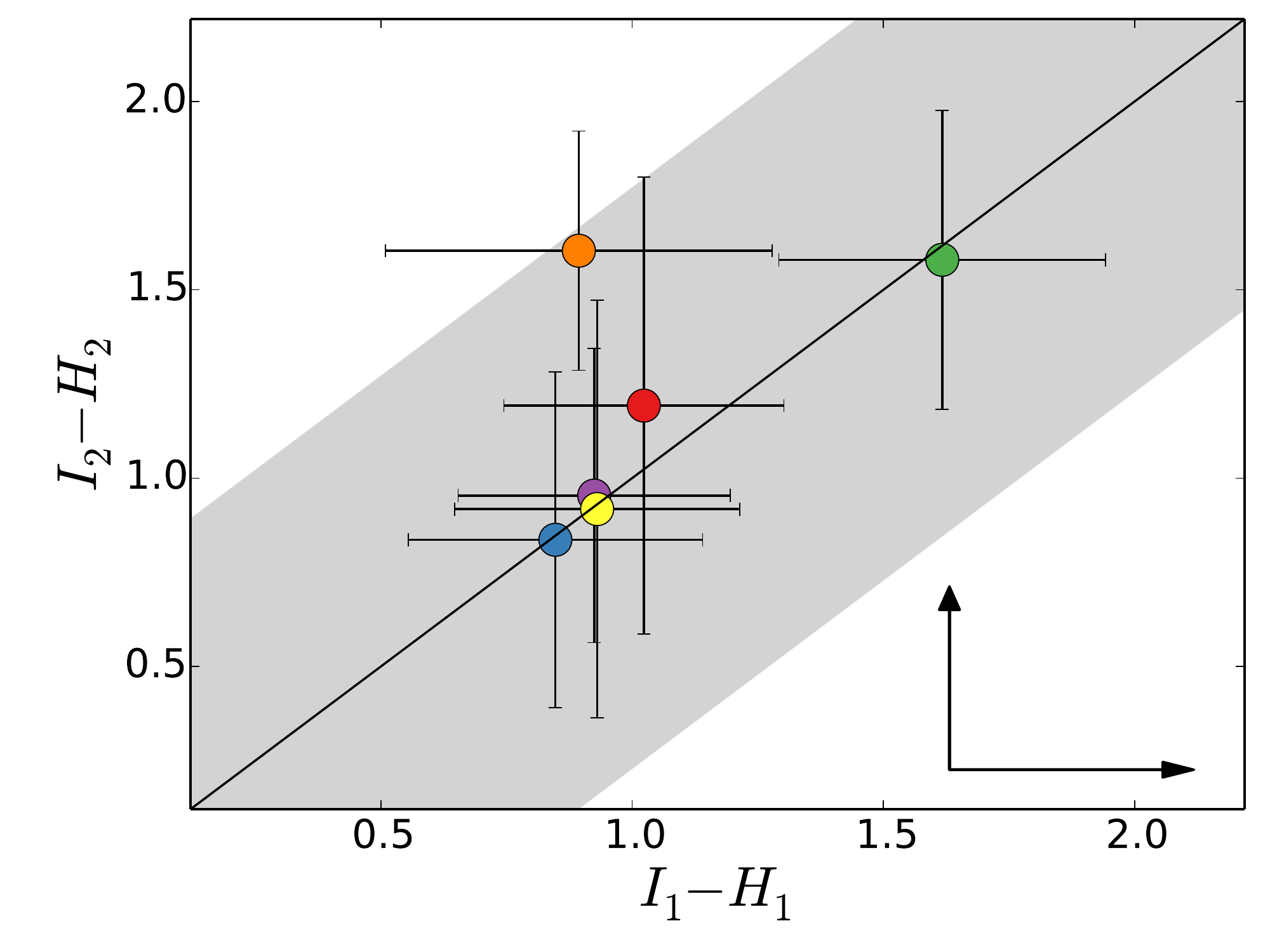} \\
\hspace*{-0.in} \includegraphics[width=0.46\textwidth]{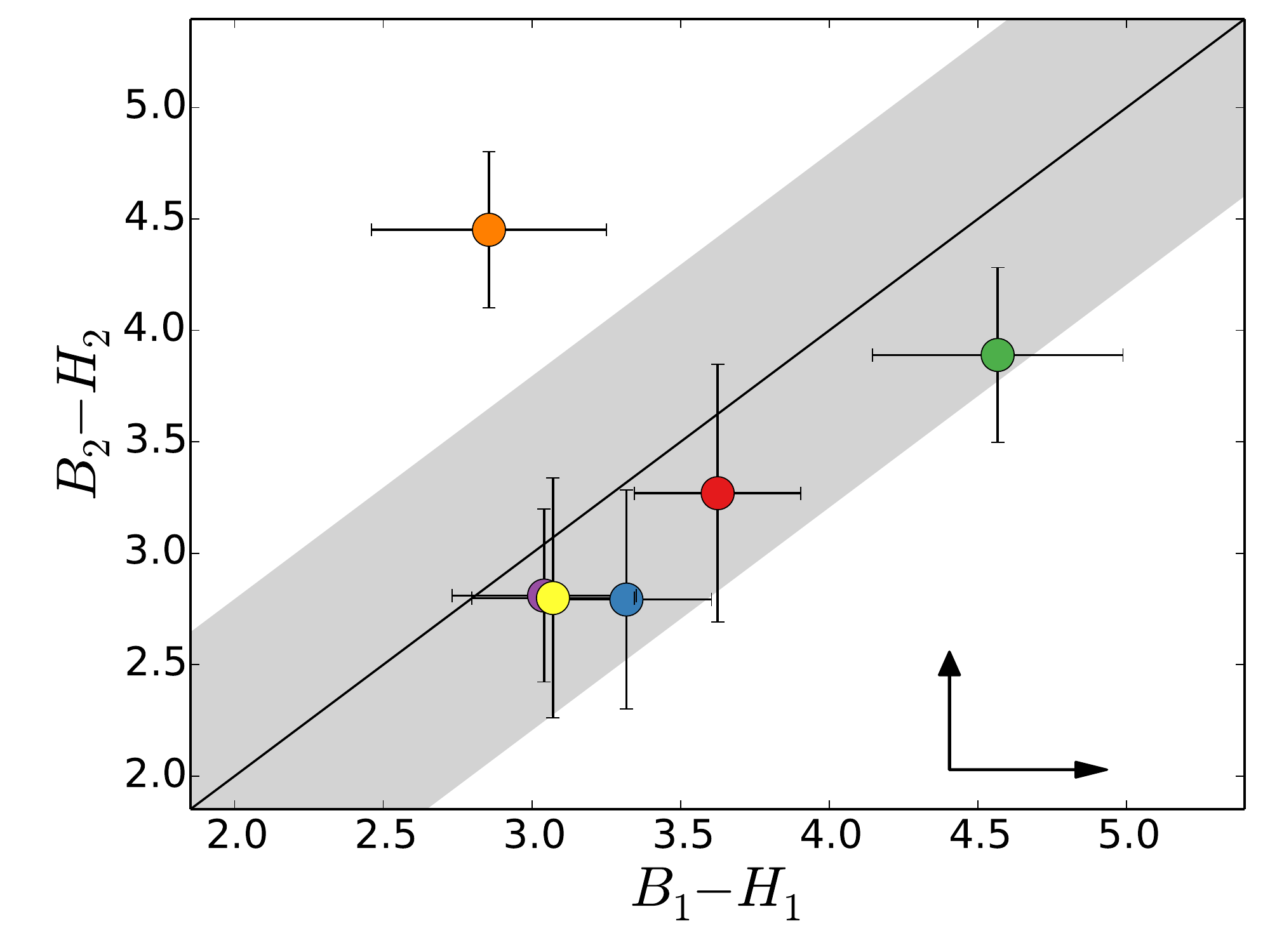}
\vspace*{-0.in}
\caption{\footnotesize{Colors in the \second~stellar core versus those in the \prim~stellar core. Top: \BItwocolor~versus \BIonecolor; middle: \IHtwocolor~versus \IHonecolor; and bottom: \BHtwocolor~versus \BHonecolor. The solid black line denotes the one-to-one relation and the grey shaded region denotes the $1\sigma$ upper and lower bounds based on the scatter and average uncertainties of the data points. Extinction arrows are shown in the lower right of each panel for the \prim~(horizontal arrow) and \second~(vertical arrow) stellar cores. The extinction arrows correspond to \nhexgal~$=10^{21}$ \uNH and are computed from the gas-to-dust relation for AGN from \citet{Maiolino:2001}, the extinction curve from \citet{Cardelli:1989}, and \Rv~$=3.1$. Note that the \second~stellar core of \offfive~is significantly reddened compared to the rest of the sample, consistent with the relatively high column density measured from the X-ray AGN at that location. 
}
}
\label{fig:plot_Hratio_Iratio_Bratio_LXratio}
\end{figure}

Figure \ref{fig:HST_IB_fiber} shows the \Ib- and \Bb-band images zoomed in on the SDSS fiber position and with the apertures used for source extraction marked. For consistency, the \Hb-band fluxes are measured from apertures based on the same procedure (marked in Figure \ref{fig:HST_H_fiber}). The \BIcolor, \IHcolor, and \BHcolor~colors for the \prim~galaxy and \second~stellar cores are listed in Table \ref{tab:gal_props} and plotted against each other in Figure \ref{fig:plot_Hratio_Iratio_Bratio_LXratio}. 

Among the four mergers, only one stellar core (the \second~stellar core of \offfive) is significantly reddened compared to the other galaxy stellar cores in the sample. The reddening is most pronounced in the \BHcolor~color (offset by $1.1\sigma$ from the sample mean) which is most sensitive to dust. These colors suggest the presence of nuclear dust that is qualitatively consistent with the potential dust lanes observed in the color composite image (Figure \ref{fig:HST_RGB}). As shown by the extinction arrows (Figure \ref{fig:plot_Hratio_Iratio_Bratio_LXratio}), the AGN in the \second~stellar core of \offfive~may be obscured by a column of $>10^{21}$ \uNH~compared to nuclei in the rest of the sample. While the red colors may in principle be due to relatively older stellar populations, we see no evidence for this scenario since the star formation rates and times since the most recent burst of star formation \citep[obtained from the SDSS spectra;][]{Thomas:2013} are within one standard deviation of the sample mean. The presence of enhanced nuclear dust compared to the rest of the sample is also reinforced by the fact that it has the largest extragalactic column density of hydrogen (\nhexgal~$=3.5\times 10^{21}$ cm$^{-2}$), measured from X-ray spectral modeling, among the sample. 

The other three mergers (\offtwo, \offfour, and \offsix) have similar colors in the \prim~and \second~stellar cores (agreement within $1\sigma$). While we do not know the relative contributions of dust, stellar continua, and photo-ionized gas from stars or AGN in these systems, the colors do not present any evidence for one of the nuclei being more obscured than the other. Therefore, no evidence is found for heavily reddened nuclei without X-ray detections in the four mergers.

The \prim~stellar core in \offfive~is the bluest of the \prim~stellar cores in the sample. The blue colors may suggest that it hosts enhanced star formation. The blue colors may also indicate that it hosts an AGN, as further suggested by the presence of two emission line systems in the SDSS fiber spectrum that are both consistent with AGN photo-ionization \citep{Ge:2012}. While the origin of the double-peaked signature is unknown since the spatial information is destroyed by the fiber, each emission line system may be associated with one of the stellar cores. The absence of an X-ray detection associated with the hypothetical AGN in the \prim~stellar core would be explained by a radiatively inefficient accretion flow and a correspondingly low Eddington ratio \citep{Ho:2009,Abramowicz:2002}. However, in this inefficient accretion state the AGN would likely not produce an optical emission line system, leaving the double-peaked spectral feature unexplained. Therefore, the AGN in the \second~stellar core may be photoionizing the NLR in the \prim~galaxy, thereby producing the blue colors. As another alternative, the AGN in the \second~stellar core may be just outside of the SDSS fiber so that its optical signature is not present in the spectrum. In this case, an X-ray faint AGN could be present in the \prim~stellar core and producing an outflow responsible for the double-peaked emission line signature and the spatially extended \Ib- and \Bb-band emission (Figure \ref{fig:HST_IB_fiber}). We consider the dual AGN hypothesis for \offfive~when discussing these galaxy mergers in the context of galaxy-SMBH co-evolution (Section \ref{sec:co-evol}).

\begin{deluxetable*}{lccccccc}
\tabletypesize{\footnotesize}
\tablecolumns{8}
\tablecaption{X-ray Source Properties.}
\tablehead{
\colhead{\name \vspace*{0.05in}} &
\colhead{\RAX} &
\colhead{\DECX} &
\colhead{\XScnts} &
\colhead{\XHcnts} &
\colhead{\Xcnts} &
\colhead{\Xdetsig} &
\colhead{\LXhard} \\
\colhead{(-) \vspace*{0.05in}} &
\colhead{(\uHMS)} &
\colhead{(\uDMS)} &
\colhead{(\uCNTS)} &
\colhead{(\uCNTS)} &
\colhead{(\uCNTS)} &
\colhead{(\uSIG)} &
\colhead{(log[\uLum])} \\
\colhead{(1)} &
\colhead{(2)} &
\colhead{(3)} &
\colhead{(4)} &
\colhead{(5)} &
\colhead{(6)} &
\colhead{(7)} &
\colhead{(8)}
}
\startdata
J0813$+$5418 \vspace*{0.05in} & 08:13:30.415 & $+$54:18:44.71 & \xrayScntsone & \xrayHcntsone & \xraycntsone & \xraysigdetone & $41.28_{-0.12}^{+0.11}$ \\
J0940$+$3113 \vspace*{0.05in} & 09:40:32.176 & $+$31:13:29.51 & \xrayScntstwo & \xrayHcntstwo & \xraycntstwo & \xraysigdettwo & $41.14_{-0.27}^{+0.18}$ \\
J1021$+$1305 \vspace*{0.05in} & 10:21:41.940 & $+$13:05:50.19 & \xrayScntsthree & \xrayHcntsthree & \xraycntsthree & \xraysigdetthree  & $41.14_{-0.14}^{+0.11}$ \\
J1114$+$4036 \vspace*{0.05in} & 11:14:58.084 & $+$40:36:12.26 & \xrayScntsfourNE & \xrayHcntsfourNE & \xraycntsfourNE & \xraysigdetfourNE  & $41.08_{-0.07}^{+0.06}$ \\
 \vspace*{0.05in} & 11:14:57.952 & $+$40:36:11.25 & \xrayScntsfourSW & \xrayHcntsfourSW & \xraycntsfourSW & \xraysigdetfourSW  & $40.97_{-0.07}^{+0.06}$ \\
J1234$+$4751 \vspace*{0.05in} & 12:34:20.247 & $+$47:51:55.84 & \xrayScntsfive & \xrayHcntsfive & \xraycntsfive & \xraysigdetfive  & $44.63_{-0.47}^{+0.71}$ \\
J2125$-$0713 & 21:25:12.470 & $-$07:13:29.88 & \xrayScntssix & \xrayHcntssix & \xraycntssix & \xraysigdetsix  & $42.51_{-0.08}^{+0.08}$
\enddata
\tablecomments{Column 1: Galaxy name; Columns 2-3: RA and DEC of the X-ray source (in the \Hb-band reference frame); Columns 4-6: soft ($0.5-2$ keV), hard ($2-10$ keV), and total ($0.5-10$ keV) source counts; Column 7: source detection significance; and Column 8: unabsorbed, rest-frame hard X-ray luminosity.}
\label{tab:xray_props}
\end{deluxetable*}

\begin{deluxetable*}{lcccccc}
\tabletypesize{\footnotesize}
\tablecolumns{7}
\tablecaption{X-ray Source Spatial Offsets.}
\tablehead{
\colhead{\name \vspace*{0.05in}} &
\colhead{\deltaANGPri} &
\colhead{\deltaSPri} &
\colhead{\PAPri} &
\colhead{\deltaANGSec} &
\colhead{\deltaSSec} &
\colhead{\PASec} \\
\colhead{(-) \vspace*{0.05in}} &
\colhead{(\uARCSEC)} &
\colhead{(\uKPC)} &
\colhead{(\uDEG)} &
\colhead{(\uARCSEC)} &
\colhead{(\uKPC)} &
\colhead{(\uDEG)} \\
\colhead{(1)} &
\colhead{(2)} &
\colhead{(3)} &
\colhead{(4)} &
\colhead{(5)} &
\colhead{(6)} &
\colhead{(7)}
}
\startdata
J0813$+$5418 \vspace*{0.05in} & $2.17\pm 0.60$ & $1.71\pm 0.47$ & 87 & $0.73\pm 0.53$ & $0.57\pm 0.42$ & 353\\
J0940$+$3113 \vspace*{0.05in} & $1.45\pm 0.71$ & $4.07\pm 2.01$ & 310 & $3.69\pm 1.36$ & $10.39\pm 3.83$ & 299\\
J1021$+$1305 \vspace*{0.05in} & $1.16\pm 0.39$ & $1.65\pm 0.55$ & 97 & $0.68\pm 0.34$ & $0.96\pm 0.49$ & 138\\
J1114$+$4036 \vspace*{0.05in} & $0.83\pm 0.43$ & $1.16\pm 0.61$ & 53 & $1.82\pm 0.71$ & $2.55\pm 1.00$ & 41\\
 \vspace*{0.05in} & $0.98\pm 0.38$ & $1.37\pm 0.53$ & 238 & $0.48\pm 0.36$ & $0.67\pm 0.51$ & 319\\
J1234$+$4751 \vspace*{0.05in} & $1.19\pm 0.46$ & $3.57\pm 1.39$ & 93 & $0.41\pm 0.28$ & $1.21\pm 0.85$ & 332\\
J2125$-$0713 & $0.19\pm 0.16$ & $0.23\pm 0.19$ & 122 & $1.33\pm 0.61$ & $1.59\pm 0.72$ & 286
\enddata
\tablecomments{Column 1: Galaxy name; Columns 2-4: angular separation, physical separation, and position angle (East of North) between the \prim~stellar core and X-ray source; and Columns 5-7: same as Columns 2-4 but between the \second~stellar core and the X-ray source.}
\label{tab:xray_offsets}
\end{deluxetable*}

\subsection{Two Ambiguous Systems}
\label{sec:ambiguous}

In the remaining two systems (\offone~and \offthree), no \second~stellar cores are detected that are interacting with the \prim~galaxy based on proximity or visual evidence of merger-related morphological disturbances in either the images or the model residuals. Moreover, the offset X-ray sources are not associated with any stellar core detections (Figure \ref{fig:HST_H_fiber}). To put upper limits on the presence of \Hb-band counterparts to the X-ray sources, we add a two-dimensional Gaussian function to the models with a peak that is constrained to be within the X-ray source centroid $1\sigma$ confidence interval. This puts upper limits of \Mtwo~$=$~\oneHtwoMgal~\Msun~and \threeHtwoMgal~\Msun~on the masses of \second~stellar cores associated with the X-ray sources in \offone~and \offthree, respectively. Applying these upper limits to the relation from \citet{Marconi:Hunt:2003} yields BH mass upper limits of \MBH~$=$~\oneHtwoMBH~\Msun~and \threeHtwoMBH~\Msun~for \offone~and \offthree, respectively. We consider the following explanations for the lack of stellar core detections associated with the X-ray sources: \\ \\
\emph{Association with the Primary Galaxy Nucleus.} The lack of any merger signatures in these systems suggests that the optically detected AGN are located in the \prim~galaxy nuclei in both cases. Therefore, since no firm stellar counterparts to the X-ray sources are detected, we first acknowledge the possibility that they are associated with the optical AGN in the \prim~galaxy nuclei. In this case, the errors on the X-ray source positions, relative to the \Hb-band \prim~galaxy nuclei, are larger than estimated. If the angular offsets of $>2''$ (\offone) and $>1''$ (\offthree) are due to uncertainty, then they would be relatively large compared to the typical \ch~absolute astrometry within $5'$ of the observation aimpoint (selection criteria for our sources; see \paperI) and therefore imply underestimated uncertainties in the X-ray source model centroids. In the specific case of \offthree, the relatively lower detection significance of the X-ray source may also suggest a spurious detection, in which case no offset X-ray source would be present. \\ \\
\emph{Background AGN.} The spatially offset X-ray sources may be associated with background AGN. This scenario would naturally explain the lack of detected stellar counterparts to the X-ray sources. We use the SDSS \mpa~galaxy catalogue \citep{Kauffmann:2003,Brinchmann:2004} and the \ch~Source Catalogue \citep{Evans:2010} to estimate surface densities and simulate the occurrence of chance projections of unrelated sources.  We find that, in the redshift and flux ranges of the \offAGNhstsztext~targets, the probability of a chance alignment between an SDSS galaxy and a \ch~source within $10''$ is $\sim10^{-4}$, corresponding to a high probability that the sources are physically related. \\ \\ 
\emph{IMBHs or Tidal Stripping.} The BH mass upper limits of \offone~and \offthree~are plausibly consistent with IMBHs of mass \MBH~$=10^{2}-10^{5}$ \Msun. These X-ray sources may be in the remnant stellar cores of tidally stripped low mass galaxies as suggested for HLX1 \citep{Farrell:2009} and other IMBH candidates \citep{King:2005,Wolter:2006,Feng:Kaaret:2009,Jonker:2010,Mezcua:2015}. However, no morphological features suggestive of past mergers are apparent in either galaxy. \\ \\
\emph{Super-Eddington Accretion and X-ray Binaries.} The luminosities of the offset X-ray sources in \offone~and \offthree~are slightly above the threshold for ultra-luminous X-ray sources (ULXs; \LXhard~$=10^{39}-10^{41}$ \uLum) and consistent with the class of hyper-luminous X-ray sources (HLXs; \LXhard~$>10^{41}$ \uLum). A possible explanation for ULXs and HLXs is super-Eddington accretion in X-ray binaries (XRBs) \citep{Colbert:Mushotzky:1999,Begelman:2002}, in which case they are predicted to occur more often in star-forming regions \citep{Madau:1998,Ghosh:White:2001,Swartz:2009}. Indeed, the X-ray source positions for both \offone~and \offthree~are coincident with $>2\sigma$ detections in the \Ib- and \Bb-band images based on the same aperture extraction used in Section \ref{sec:dual_agn}. These detections may be associated with stars or photo-ionized gas in star-forming regions. Moreover, their relatively red colors (Figure \ref{fig:plot_Hratio_Iratio_Bratio_LXratio}) may be associated with dust that is responsible for obscuring the X-ray signature of the central AGN. Therefore, if the X-ray sources are not associated with the optical AGN in the \prim~galaxy nucleus, we consider the XRB interpretation to be the most likely explanation.

\begin{figure*}[t!] $
\begin{array}{c c}
\hspace*{-0.in} \includegraphics[width=0.49\textwidth]{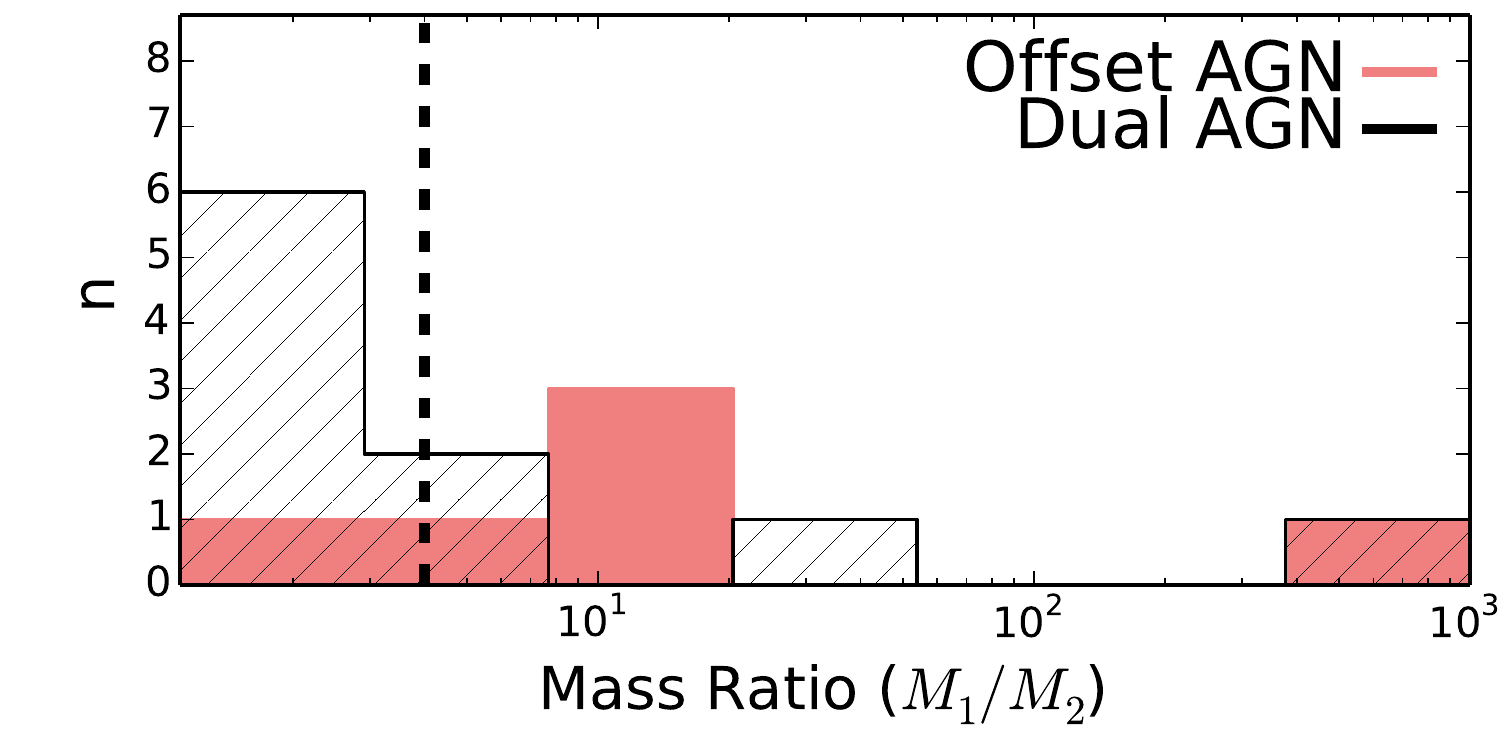}
\hspace*{-0.in} \includegraphics[width=0.49\textwidth]{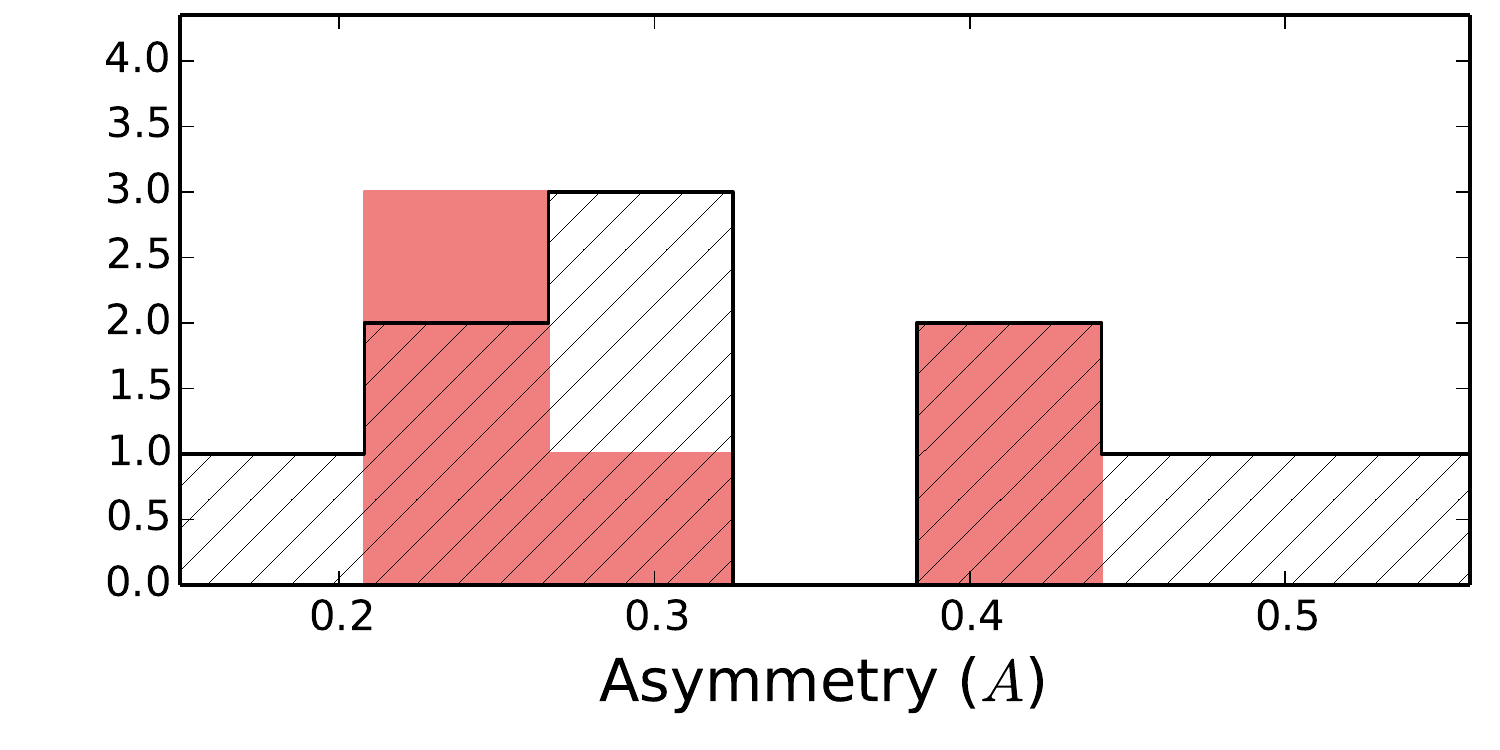}
\end{array} $
\caption{\footnotesize{Left: distribution of mass ratios between the \prim~and \second~stellar cores (\mratio); right: distribution of asymmetry indices (\asymm). In both panels the offset AGN values are denoted by the filled, light red histogram, and the dual AGN values are denoted by the hatched, black histogram. In the left panel, the black, dashed line denotes \mratio~$=4:1$, our adopted division between major and minor mergers. Note that the offset AGN are predominantly in minor mergers and generally have larger mass ratios compared to the dual AGN. On the other hand, the asymmetry distributions are similar among both samples.}
}
\label{fig:hist_mratio_asymm}
\end{figure*}

\section{Spatially Offset X-Ray Sources as a Selector of AGN in Galaxy Mergers}
\label{sec:selector}

Selection of galaxy mergers hosting AGN is vital for thorough studies of galaxy-SMBH co-evolution. While large scale galaxy pairs in early merger stages are easily identifiable, angular resolution limits make more advanced mergers particularly difficult to find. However, late-stage galaxy mergers can, in principle, be identified based on signatures of spatially offset X-ray sources. Moreover, since the selection is not based on visual inspection or asymmetries, it is not biased toward major mergers with large morphological disturbances. 

The \hst~imaging presented in Section \ref{sec:nature_xray} indicates that $2/6$ of the spatially offset X-ray sources are direct signatures of mergers since they are associated with \second~galaxies that are interacting with the \prim~galaxy. Furthermore, one of them is a minor merger with no discernible morphological disturbances in the SDSS imaging and therefore would likely be missed by selections based on visual or asymmetry signatures. Another $2/6$ of the spatially offset X-ray sources are in mergers but the most likely NIR counterparts are the \prim~stellar cores. Therefore, the spatially offset X-ray selection may be related to discrepancies in the astrometric solutions between the \ch-SDSS registrations and the \ch-\hst~registrations. However, we also note that few sources are available for matching within the \hst/F160W FOV and hence the astrometric uncertainties are larger in both cases compared to the original selection.

Previous methods of selecting late-stage galaxy merger candidates based on spectroscopic evidence in the form of double-peaked emission lines yielded a merger selection rate of $\sim10\%$ \citep{Shen:2011b,Fu:2012,Mueller-Sanchez:2015,Nevin:2016}. More recent selection based on velocity offset emission lines \citep{Comerford:Greene:2014} may yield an even lower merger selection rate \citep{Comerford:2017a}. While the sample size of \offAGNhstsztext~galaxies here is small, the direct merger selection of $33\%$ may indicate that spatially offset AGN are a more efficient merger selector and could be attributed to the focus on hard X-ray signatures that are produced near the SMBH accretion disk, thereby removing the effects of spatially extended features related to inflows or outflows. Furthermore, the merger selection efficiency of spatially offset AGN may be even higher when restricted to the most luminous X-ray AGN. In particular, when excluding the two sources with the lowest hard X-ray luminosities (\offone~and \offthree), the fraction of directly selected mergers increases to $2/4$ ($50\%$).

Selection of spatially offset X-ray sources can be readily applied to vast samples of archival data to yield large numbers of AGN in candidate galaxy mergers. Offset X-ray sources can also reveal candidate recoiling SMBHs that are the eventual products of some galaxy mergers \citep{Kim:2017}. In addition to the use of \ch~data, recent studies have utilized archival radio imaging to find several offset AGN \citep{Condon:2017} or candidate offset AGN \citep{Makarov:2017,Skipper:2018}.

Finally, as demonstrated by our sample, this technique allows for the selection of spatially offset X-ray sources out to intermediate redshifts. This can aid in the search for ULX and HLX candidates (which in previous studies has been mainly focused on nearby galaxies) by significantly increasing ULX/HLX sample sizes and permitting tests of how their environments evolve. Moreover, this technique will be even more effective when used in conjunction with the increased sensitivity of future next generation X-ray observatories such as \lynx.

\section{Offset AGN in the Context of Galaxy-SMBH Co-Evolution}
\label{sec:co-evol}

The mergers that we identify have physical separations of $<10$ kpc, with the smallest separation being just over $1$ kpc. Moreover, three-fourths of the systems are minor mergers that are difficult to find by most merger selection techniques. Therefore, we use the merger sample for comparisons between AGN in minor and major mergers and to provide unique insight about merger-driven triggering of AGN at small separations where theory predicts it will peak.

We augment the sample with the full subset of \offAGNsdsssztext~offset AGN from \paperI~without \hst~imaging for which we are able to resolve the individual stellar cores associated with the merger using SDSS imaging. In these cases, the X-ray AGN is consistent with being at the galaxy central stellar core and offset from a secondary stellar core that is present outside of the fiber but within $20$ kpc in projected separation and at a similar redshift (see \paperI~for details). For the \offAGNsdsssztext~offset AGN with only \sdss~imaging, the mass ratios are calculated using the same procedure used for the offset AGN with \hst~imaging.

\subsection{Formation of Offset AGN Versus Dual AGN}
\label{sec:form_off_dual}

The mechanisms that drive fuel to the nuclear regions of galaxies during mergers can also determine whether or not one AGN (offset AGN) or two AGN (dual AGN) are triggered. In particular, a basic prediction is that the more efficiently gas and dust is transported to the nuclear regions the more likely a dual AGN will form rather than an offset AGN. An increasing body of theoretical work now shows that the parameters most strongly in control of offset AGN and dual AGN triggering are the merger mass ratio, nuclear separation, and overall gas mass or gas fraction \citep{Van_Wassenhove:2012,Blecha:2013,Capelo:2015,Steinborn:2016,Rosas-Guevara:2018,Steinborn:2018}. Understanding the scenarios leading to the formation of offset AGN versus dual AGN will put constraints on the physical mechanisms that build up SMBH mass in galaxy mergers and will inform cosmological models of hierarchical galaxy-SMBH co-evolution. Therefore, we test these predictions by comparing properties of the offset AGN to those of a comparison dual AGN sample.

The dual AGN sample consists of \dualAGNtotsztext~galaxy mergers hosting AGN. Of these \dualAGNtotsztext~dual AGN, \dualAGNCsztext~are from \citet{comerford:2015}, while the remaining \dualAGNLsztext~are the mergers from \citet{Liu:2013} for which mass ratios have been measured \citep{Shangguan:2016}. Similar to the offset AGN, these systems were originally selected from the \sdss~to have optical emission lines consistent with AGN photo-ionization \citep{Kewley:2006}. Moreover, each has also been imaged by \hst~in a \wfcthree~NIR filter and observed by \ch. However, an important distinction is that the emission lines of the dual AGN were selected to have double-peaked profiles that suggest the possible orbital motion of two AGN within a galaxy merger. Follow-up optical longslit spectroscopy at orthogonal position angles reveals that each emission line system is a spatially distinct AGN-photoionized narrow line region (NLR). Both NLRs of each galaxy merger are spatially coincident with an \hst~NIR stellar core hosting a SMBH, suggesting the presence of a dual AGN. From here on we also include \offfour~in the dual AGN sample since our analysis finds it to be a candidate dual AGN (Section \ref{sec:loc_agn}). This leaves final samples of \offAGNplotsztext~offset AGN and \dualAGNplotsztext~dual AGN used in the following analyses.

\subsubsection{The Role of Merger Morphology}
\label{sec:gal_morph}

Simulations of SMBHs in evolving galaxy mergers generally predict that dual activation is more frequent in major mergers \citep{Van_Wassenhove:2012,Blecha:2013,Capelo:2015}. Indeed, the observational results from \citet{Comerford:2018} have recently shown that dual AGN are predominately in major mergers. Additionally, recent simulations have specifically predicted that offset AGN are more often found in minor mergers \citep{Steinborn:2016}. We test these predictions of offset AGN morphology by comparing the mass ratio distribution of the offset AGN versus that of the dual AGN in the left panel of Figure \ref{fig:hist_mratio_asymm}.

The offset AGN have a mean \mratio~value of \OffMRATIOmean, while the dual AGN have a mean \mratio~value of \DualMRATIOmean. The difference between these mean values is at the \OffDualMRATIOmeandiffsig~level, and a Kolmogorov-Smirnov (KS) test yields a statistic of \OffDualMRATIOKSstat~and null hypothesis probability of \OffDualMRATIOKSprob~that the two samples are drawn from different distributions. When applying the major versus minor merger threshold of \mratio~$=$~\MRATIOthresh, we find that $83^{+16}_{-47}\%$ of the offset AGN are in minor mergers while only $30^{+35}_{-23}\%$ of the dual AGN are in minor mergers. This difference between the minor merger fractions of the offset AGN and dual AGN is strengthened if \offfive~is also considered to be a dual AGN (see Section \ref{sec:dual_agn}), with \OffexcMAJORNUMfour~major mergers and \OffexcMINORNUMfour~minor mergers in the offset AGN sample compared to $8$ major mergers and $3$ minor mergers in the dual AGN sample. Thus, the offset AGN are preferentially found in minor mergers whereas the dual AGN show weaker evidence for a preference (though with the majority of mass ratios corresponding to major mergers).

To understand the offset AGN and dual AGN mass ratios in the overall context of galaxy mergers, we compare our results to estimates of minor and major merger fractions for galaxies selected independently of AGN signatures. Using the \citet{Lotz:2011} redshift-dependent merger fraction function and averaging over the redshift range of our sample, we find minor and major merger fractions in the general galaxy population of \MINORFRACZAVGLOTZ~and \MAJORFRACZAVGLOTZ, respectively. This comparison shows that the offset AGN minor and major merger fractions are consistent with those of galaxies without AGN. By implication, selection of galaxy mergers with a single AGN does not preferentially target major mergers when compared to the general population of galaxy mergers at the same redshifts. Since minor mergers are far more common than major mergers, offset AGN may therefore represent an important component of SMBH growth that occurs in mergers. On the other hand, the dual AGN minor (major) merger fractions are (lower) higher than those of non-AGN galaxies. This implies that selection of systems with dual AGN does preferentially target major mergers when compared to the general population of galaxy mergers at the same redshifts.

We also compare the asymmetries between the offset AGN and dual AGN in the right panel of Figure \ref{fig:hist_mratio_asymm}. Asymmetries are measured as described in \citet{Barrows:2017b} and based on the procedure outlined in \citet{Conselice:2000}. The measurements are made by first creating a galaxy image that is rotated $180^{\circ}$ about the center (chosen to be the \prim~stellar core location). Then the rotated image is subtracted from the original image and the difference is normalized by the original image sum to yield the source asymmetry parameter. The sky asymmetry is then measured and subtracted from the source asymmetry to produce the final asymmetry value, \asymm. The images used for the asymmetry measurements are the same as those used for the mass ratio measurements. We find no evidence that the two samples are drawn from different distributions (KS statistic of \OffDualASYMMKSTstat~and null hypothesis probability of \OffDualASYMMKSTprob). This result may suggest that the overall morphological disturbance at large radii has little effect on whether or not one or two AGN are triggered in a merger. Rather, the more important effect is the gravitational force exerted by the compact stellar cores on the surrounding gas and dust.

\begin{figure}[t!]
\hspace*{-0.05in} \includegraphics[width=0.49\textwidth]{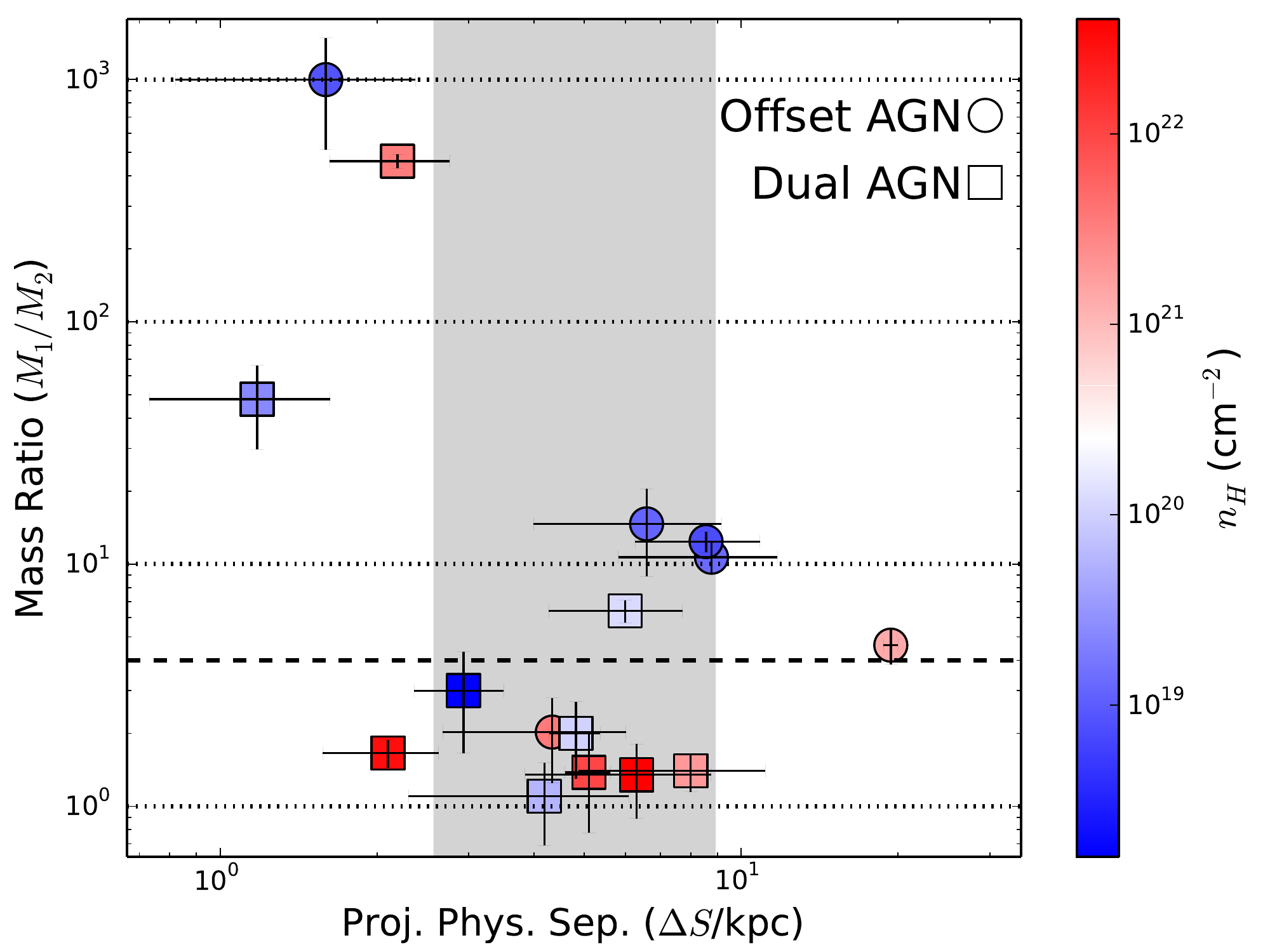}
\vspace*{-0.in}
\caption{\footnotesize{Mass ratio (\mratio) as a function of projected nuclear physical separation (\deltas) for the offset AGN (circles) and dual AGN (squares). The points are color-coded by extra-galactic column density (\nhexgal) measured from X-ray spectral fits (\paperI). The horizontal dashed line represents the demarcation between minor and major mergers, and the horizontal dotted lines denote order-of-magnitude steps in mass ratio. The grey-shaded region denotes the standard deviation of the projected physical separations about the mean. Note that the dual AGN in minor mergers have relatively small separations and potentially higher gas masses (as probed by \nhexgal) compared to offset AGN.}
}
\label{fig:mratio_physep}
\end{figure}

\subsubsection{The Combined Effects of Mass Ratio, Physical Separation, and Gas Mass}
\label{sec:roles_sep_gas}

The results from Section \ref{sec:gal_morph} show that mass ratio may play a significant role in the triggering of one versus two AGN in a merger. However, numerical simulations also find a strong dependence on merger stage. In particular, they predict that the frequency of offset AGN exceeds that of dual AGN at large pair separations, but that dual AGN become more common at increasingly smaller pair separations \citep{Van_Wassenhove:2012,Blecha:2013,Capelo:2015,Steinborn:2016}. This suggests an evolutionary transition from offset AGN to dual AGN as a merger progresses toward later stages. However, the relative importance of mass ratio and merger stage on the formation of offset AGN versus dual AGN has not yet been tested observationally.

\begin{figure}[t!]
\hspace*{-0.1in} \includegraphics[width=3.55in]{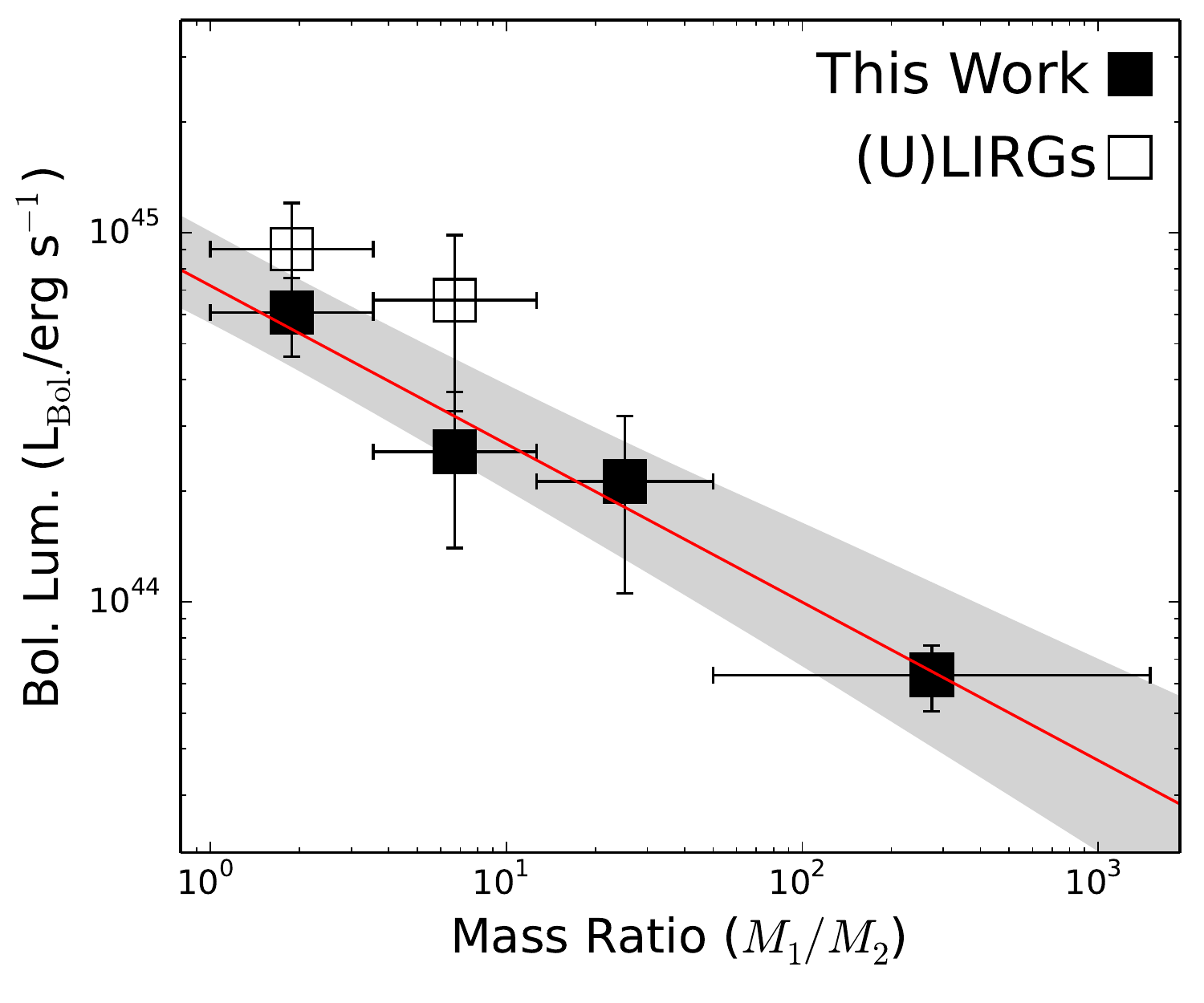}
\vspace*{-0.15in}
\caption{\footnotesize{AGN bolometric luminosity (\LBol) as a function of the mass ratio between the primary and secondary stellar core (\mratio). The data points are binned along the X-axis using the Freedman-Diaconis rule and adjusted to include $\geq 2$ points per bin. The filled data points represent the sample of optically selected and X-ray detected offset AGN and dual AGN analyzed in this work. The solid, red lines indicate the error-weighted best-fit power-law functions to the filled data points, while the light, grey shaded regions indicate the lower and upper $68.3\%$ confidence bounds. The open data points represent the comparison sample of (U)LIRGs. For all data points the X-axis error bars represent the bin width, and the Y-axis error bars are the average Y-value uncertainty in each bin. We note that the general trend is consistent with AGN accretion rates increasing toward major mergers and hence toward greater tidal disturbances.}
}
\label{fig:plot_LBolB_fEddB_MRATIO}
\end{figure}

Therefore, in Figure \ref{fig:mratio_physep} we use our sample to show the bivariate distributions of mass ratios and nuclear separations and to examine their combined effects on offset AGN versus dual AGN triggering. Interestingly, Figure \ref{fig:mratio_physep} shows that the dual AGN with the largest mass ratios (\mratio~$>10$) are also among the pairs with the smallest physical separations of the merger sample (offset by $>1\sigma$ relative to the mean). While this result is small in significance, it may suggest that triggering dual AGN in mergers with large mass ratios (and hence in less morphologically disturbed systems) requires advanced merger stages when a sufficient amount of fuel has migrated to the nuclei.

To further test the dependence on the fuel supply, we have color coded each point by the value of its extragalactic column density (\nhexgal) measured from X-ray spectral fits (\paperI). Values of \nhexgal~are a measure of hydrogen in the host galaxy along the line of sight to the X-ray source and can be used to assess the level of gas near the AGN. Among the largest mass ratios of the sample, the dual AGN with the largest mass ratio (\mratio~$>100$) has the largest extragalactic column density, particularly when compared to offset AGN with similar mass ratios. A possible interpretation is that, for large mass ratios and small separations, triggering a dual AGN versus an offset AGN may require high gas masses. This result is qualitatively consistent with numerical results from \citet{Steinborn:2016} who find that, compared to offset AGN, dual AGN are generally in more gas rich systems that will lead to triggered accretion onto both SMBHs.  This expectation is corroborated by the more than order-of-magnitude difference in the mean \nhexgal~for the dual AGN (\Dualnhmean~\uNH) compared to the offset AGN (\Offnhmean~\uNH).

While the statistical significance of these combined results is small, they are consistent with the theoretical picture in which offset AGN and dual AGN formation is tied to mass ratio, physical separation, and nuclear gas mass. Moreover, these hints point to the potential power of large offset and dual AGN samples for understanding the detailed physical mechanisms of SMBH growth in mergers.

\subsection{Merger-Driven SMBH Growth}
\label{sec:merger_growth}

A long history of numerical simulations predicts that major galaxy mergers can be efficient mechanisms for driving gas and dust to the nuclear regions of galaxies and hence for fueling AGN \citep{Barnes:1991,Hopkins:2009c}. Indeed, the results from Section \ref{sec:roles_sep_gas} demonstrate that the merger parameters of mass ratio and nuclear separation may play roles in driving fuel to both SMBHs versus only one. However, whether mergers can significantly elevate the accretion rates onto SMBHs or if the triggered accretion rates are instead similar to those of AGN in non-mergers is unclear. For example, the recent analytical results from \citet{Weigel:2018} suggest that the observed AGN major merger fraction can be satisfactorily produced by a distribution of Eddington ratios with the same shape as the general AGN population. Their result implies that, while major mergers build more gas-rich systems, the rate at which gas falls to the nuclear regions is similar to non-mergers.

On the other hand, the generally larger column densities of the dual AGN, compared to the offset AGN, seen in Section \ref{sec:roles_sep_gas} suggest that major mergers may actually supply the SMBHs with enhanced fuel supplies relative to minor mergers. However, few studies have investigated the role of minor mergers for AGN triggering and the effect of mass ratio on SMBH growth has not yet been investigated down to the minor merger regime and $<10$ kpc separations. Therefore, we use our sample of offset and dual AGN to examine how AGN bolometric luminosity evolves over three orders of magnitude in mass ratio during late-stage mergers.

We calculate bolometric luminosities (\LBol) from the extinction-corrected, integrated \oiii~luminosities measured in the SDSS fiber spectra and the bolometric correction from \citet{Trump:2015} based on the sample of \citet{Lamastra:2009}. We note that, while hard X-ray luminosities are in principle a more direct tracer of AGN radiative output, the samples from \citet{comerford:2015} and \citet{Liu:2013} both suggest that nuclear obscuration may introduce significant uncertainties in the X-ray spectral modeling.

Figure \ref{fig:plot_LBolB_fEddB_MRATIO} shows \LBol~as a function of \mratio. The data points have been binned by mass ratio and each bin includes $\geq 2$ sources. The confidence intervals around the best-fitting power-law functions are determined by adding simulated random uncertainties (assuming the errors follow a normal distribution) and refitting until the upper and lower uncertainties converge. The best-fit power-law function relating \LBol~and \mratio~has a negative slope that is offset from zero at a significance of \OffDualLBolLOIIIMRATIOfracsig. This result indicates that more quickly growing SMBHs are in galaxy mergers with mass ratios close to unity and that the effect of mass ratio persists down to small pair separations. Furthermore, a correlation between mass ratio and bolometric luminosity suggests that information about the merger dynamics is preserved in the gas that is ultimately accreted onto the SMBHs. Indeed, the bolometric luminosities of our sample are quantitatively consistent with the simulations from \citet{Steinborn:2016} who predict that the merger mass ratio does have a strong effect on accretion rates of merger-driven SMBH growth.

We acknowledge that these results only apply to AGN that are optically selected and detected at X-ray energies of $<10$ keV, thereby introducing a bias toward relatively unobscured systems. Therefore, for comparison we also show in Figure \ref{fig:plot_LBolB_fEddB_MRATIO} a sample of (Ultra-)Luminous Infrared Galaxies, or (U)LIRGs, hosting AGN with column densities often approaching Compton-thick levels \citep{Komossa2003,Bianchi2008,Piconcelli2010}. These (U)LIRGs each host an AGN detected by the \swift~\batname~(\bat) survey, and they are a subset of the mergers from \citet{Koss:2012} with pair separations of $<20$ kpc as in our sample. The bolometric luminosities are determined from the \bat~ultra-hard ($14-195$ keV) X-ray luminosities and the bolometric correction of \citet{Vasudevan:2009}.

Figure \ref{fig:plot_LBolB_fEddB_MRATIO} shows that the (U)LIRGs are dominated by mass ratios of \mratio$<10$ and with average bolometric luminosities that are larger than for our optically selected and X-ray detected sample. Since (U)LIRGs are extremely gas rich and dusty \citep{Sanders:1988a,Sanders:1988,Canalizo:2001}, this comparison is consistent with the expectation that enhancements in merger-driven AGN triggering are coincident with large reservoirs of material for accretion. However, the bolometric luminosities of the (U)LIRGs are only larger than our sample by $\lesssim1\sigma$. This small enhancement suggests that, while the presence of gas and dust is a requirement for AGN triggering in any context, the rate at which gas and dust is transported to the nuclei in mergers is sensitive to the mass ratio and highest in major mergers.

\section{Conclusions}
\label{sec:conclusions}

We analyze new \hst~imaging for a sample of \offAGNhstsztext~galaxies selected to host X-ray sources that are spatially offset from the galaxy nucleus. We use the \hst~imaging to constrain the merger scenario for each host galaxy and determine the likely nature of the X-ray sources. This analysis yields the following conclusions:

\begin{itemize}

\item Four out of the \offAGNhstsztext~X-ray sources are in galaxy mergers, and the nuclear separations are $<10$ kpc in all cases. In two of these mergers the X-ray source is associated with the \prim~galaxy, and in the other two mergers it is associated with a spatially offset smaller, companion galaxy. 

\item The host galaxies for two of the \offAGNhstsztext~X-ray sources do not show any evidence of mergers. These X-ray sources may be associated with the optically selected AGN in the \prim~galaxy nucleus or may be ULXs/HLXs produced by super-Eddington accretion in XRBs. The second possibility suggests the potential for systematic searches of offset X-ray sources to identify large populations of ULXs and HLXs out to intermediate redshifts.

\end{itemize}

After combining this sample with \offAGNsdsssztext~additional offset AGN from our selection procedure for which the merger mass ratios are known, we specifically compare the properties of offset AGN to those of dual AGN. The dual AGN have similar selection criteria to the offset AGN and known mass ratios. Our aim is to understand the physics of single AGN triggering versus dual AGN triggering and how galaxy properties relate to merger-driven SMBH growth. The dynamic range of mass ratios (\fivemratio~$-$~\sixmratio) and small physical separations (\fourphysep~$-$~\twophysep~kpc) make this sample ideal for examining these connections over a wide range of tidal disturbances and during advanced merger stages when SMBH growth is predicted to peak. Our conclusions are as follows:

\begin{itemize}

\item Galaxy mergers with only a single AGN are predominantly found in minor mergers. This observation corroborates predictions from simulations that mergers with only one AGN are preferentially those with large mass ratios. Moreover, the mass ratios among single AGN are similar to the overall population of galaxy mergers, implying that mergers triggering only one AGN are not preferentially found in more disturbed galaxies compared to mergers without AGN.

\item Relative to offset AGN, dual AGN show a preference toward major mergers and larger nuclear gas masses. This result suggests that the merger mass ratio, and corresponding tidal forces, has a strong effect on the number of AGN triggered in a merger while also supporting the hypothesis that dual AGN triggering requires enhanced supplies of nuclear fuel for accretion.

\item The AGN bolometric luminosities increase toward smaller mass ratios (major mergers). This result suggests that merger morphology affects the level of SMBH accretion. Moreover, for small mass ratios our sample has accretion levels that are statistically consistent with a comparison sample of gas-rich (U)LIRGs, an indication that even mergers with enhanced reservoirs of fuel require significant tidal disturbances to drive SMBH accretion.

\end{itemize}

Finally, our results suggest that spatially offset X-ray sources may be an effective method for identifying AGN in late-stage galaxy mergers and ULXs/HLXs at intermediate redshifts from archival imaging data alone. The same principle technique can also be applied to radio imaging since it can achieve the requisite spatial resolution and is sensitive to radio loud AGN. With large volumes of archival X-ray and radio imaging becoming increasingly more available, offset AGN are a potentially promising means of studying AGN triggering in galaxy mergers.

The authors thank an anonymous referee for a detailed and constructive report that greatly improved the quality of the paper. The results reported here are based on observations made with the NASA/ESA \emph{Hubble Space Telescope}, obtained at the Space Telescope Science Institute, which is operated by the Association of Universities for Research in Astronomy, Inc., under NASA contract NAS 5-26555. These observations are associated with program number GO-14068. Support for this work was also provided by NASA through \ch~Award Number AR5-16010A issued by the \emph{Chandra X-ray Observatory Center}, which is operated by the Smithsonian Astrophysical Observatory for and on behalf of NASA under contract NAS8-03060. The scientific results reported in this article are based in part on observations made by the \emph{Chandra X-ray Observatory}, and this research has made use of software provided by the \ch~X-ray Center in the application packages \ciao~and \sherpa.

\end{document}